\documentclass[10pt, journal]{IEEEtran}
\usepackage{cite}
\usepackage{amsmath,amssymb,amsfonts}
\usepackage{algorithmic}
\usepackage[dvipdfmx]{graphicx}
\usepackage{textcomp}

\usepackage{revision}

\def\BibTeX{{\rm B\kern-.05em{\sc i\kern-.025em b}\kern-.08em
    T\kern-.1667em\lower.7ex\hbox{E}\kern-.125emX}}
\begin{document}

\title{Combining IOTA and Attribute-Based Encryption for Access Control in the Internet of Things}

\author{
	Ruka Nakanishi,
	Yuanyu Zhang, \IEEEmembership{Member, IEEE},
	Masahiro Sasabe, \IEEEmembership{Member, IEEE},
	and Shoji Kasahara, \IEEEmembership{Member, IEEE}
\thanks{
This paper is an extended version of \cite{nakanishi2020iota}, which was presented at the 2nd Conference on Blockchain Research and Applications for Innovative Networks and Services (BRAINS) in 2020. This work was supported in part by the Japan Society for the Promotion of Science (JSPS) KAKENHI (A) under Grant 19H01103 and the Support Center for Advanced Telecommunications (SCAT) Technology Research Foundation. The corresponding author is \emph{Y. Zhang (yy90zhang@ieee.org)}.

R. Nakanishi, Y. Zhang, M. Sasabe and S. Kasahara are with the Graduate School of Science and Technology, Nara Institute of Science and Technology, 8916-5 Takayama, Ikoma, Nara, 630-0192, Japan. Email: nakanishi.ruka.nm0@is.naist.jp, \{yy90zhang, m-sasabe, kasahara\}@ieee.org.}
}

\maketitle

\IEEEpubidadjcol
\bstctlcite{IEEEexample:BSTcontrol}

\begin{abstract}
Unauthorized resource access represents a typical security threat in the Internet of things (IoT), while
distributed ledger technologies (e.g., blockchain and IOTA) hold great promise to address this threat.
Although blockchain-based IoT access control schemes have been the most popular ones,
they suffer from several significant limitations, such as high monetary cost and low throughput of processing access requests.
To overcome these limitations, this paper proposes a novel IoT access control scheme by combining the fee-less IOTA technology and the Ciphertext-Policy Attribute-Based Encryption (CP-ABE) technology.
To control the access to a resource, a token, which records access permissions to this resource, is encrypted by the CP-ABE technology and uploaded to the IOTA Tangle (i.e., the underlying database of IOTA).
Any user can fetch the encrypted token from the Tangle, while only those who can decrypt this token are authorized to access the resource.
In this way, the proposed scheme enables not only distributed, fee-less and scalable access control thanks to the IOTA but also fine-grained attribute-based access control thanks to the CP-ABE. We show the feasibility of our scheme by implementing a proof-of-concept prototype system
and evaluate its performance in terms of access request processing throughput.
\end{abstract}


\section{Introduction}\label{sec:introduction}
\pagenumbering{arabic}
By virtue of the rapid advancement in the Internet of Things (IoT),
an unprecedented number of devices are connected to the Internet nowadays, from sensors to home appliances, as well as automobiles \cite{gartner}.
The tremendous amount of data these devices collect from the physical world has opened up new opportunities for smart applications
in various fields including logistics, healthcare and manufacturing \cite{ben2019internet, qadri2020future, yang2019internet}.

On the other hand, IoT devices are vulnerable to security attacks represented by unauthorized access
\cite{haddadpajouh2019survey, ande2020internet, butun2019security, attack},
because some devices, especially sensors and actuators, are not equipped with enough security measures due to their limited computing power.
Since IoT devices often handle personal information and play an important role in various task executions,
unauthorized access to them can result in serious issues such as information leakage and malfunction, greatly threatening our safety and privacy.
Therefore, protecting the devices by enforcing appropriate \emph{access control} is essential in the IoT era.

Access control is the process of defining the authority of users (i.e., subjects)
and making sure that only authorized subjects can access the objects (e.g., devices and data).
In order to enforce access control, the access rights of subjects, i.e., information about which subject can access which resource,
has to be recorded in some form and referred to when judging if a given subject is eligible to access some resource.
In most of the access control systems widely used nowadays, such information is stored in a centralized server in order to facilitate management.
However, this leads to various limitations especially in large-scale environments like the IoT, such as a single point of failure, performance bottleneck and vulnerability against tampering by malicious users.

In order to enforce reliable and distributed access control, various access control schemes applying the blockchain technology have been proposed recently
\cite{
xu2018blendcac, xu2019exploration, nakamura2019capbac, nakamura2020exploiting,
dukkipati2018decentralized, maesa2019blockchain, yutaka2019abac, zhang2020attribute,
cruz2018rbac, rahman2020context,
zhang2019IoT, sultana2020data, novo2018blockchain, ouaddah2016fairaccess,
maesa2017blockchain, pinno2017controlchain, ding2019novel, zhu2018tbac
}.
Blockchain is the underlying technology of cryptocurrencies such as Bitcoin \cite{btc} and Ethereum \cite{eth}, which is a distributed database
that is kept cooperatively in Peer-to-Peer (P2P) networks. Peers sync, verify and reach consensus on the data to record to the blockchain,
making the database tamper resistant. In addition, the data are distributed since every peer keeps its own copy of the entire blockchain.
Owing to these attractive properties, blockchains are considered suitable for storing access rights.

Another promising feature in the blockchain technology are smart contracts, which are executable codes that are embedded in the blockchain.
Implemented in some blockchains including Ethereum \cite{ethsmartcontract},
smart contracts transform the blockchain system from a mere database to a distributed and trustworthy computing platform.
In terms of access control, this can be used to perform computations like managing access rights and processing access requests,
which has led to various smart-contract-based access control schemes for the IoT
\cite{
xu2018blendcac, xu2019exploration, nakamura2019capbac, nakamura2020exploiting,
dukkipati2018decentralized, maesa2019blockchain, yutaka2019abac, zhang2020attribute,
cruz2018rbac, rahman2020context,
zhang2019IoT, sultana2020data, novo2018blockchain, ouaddah2016fairaccess
}.

Although blockchain-based access control schemes have overcome the limitations in conventional access control schemes,
they have also given rise to two main drawbacks deriving from the underlying blockchains.
First, they incur monetary cost to users, since users need to pay some fee to the peers who manage and update the blockchain \cite{nist2018blockchain},
i.e., verification of the validity of the access rights/policies stored on the blockchain and execution of smart contracts for processing access requests.
Second, they suffer from low throughput of processing the access requests.
In blockchain, newly incoming data are never considered valid unless they are verified, included in a block and the block has been appended to the blockchain.
Although new blocks are added to the blockchain at regular intervals, each block can contain only a limited amount of data,
lowering the throughput when there is an increased amount of incoming data \cite{conoscenti2016blockchain}.
These two shortcomings have to be addressed in order to adapt to large-scale IoT systems.

To address the limitations in blockchain-based schemes,
there has been an attempt to apply the IOTA technology \cite{iota} to access control recently.
IOTA is a next-generation distributed ledger technology designed for the IoT.
It aims to solve the limitations that blockchains face, i.e., low throughput and high transaction fee,
which are addressed by adopting a different data structure for the ledger and changing the way consensus is reached, respectively.
These features will be introduced in greater detail in Section \ref{sec:preliminaries}.
Based on IOTA, an access control scheme called the Decentralized Capability-based Access Control framework using IOTA (DCACI) has been proposed in \cite{dcaci}.
In the scheme, the distributed ledger of IOTA, called the Tangle, is used to store the subjects' access rights in the form of tokens.
This enables the access rights to be stored in a distributed and tamper-resistant fashion, similar to blockchain-based schemes.
In addition to this, DCACI achieves fee-less access control with high throughput thanks to IOTA.
However, the DCACI scheme faces some limitations in terms of scalability and security, like heavy token management and the lack of security considerations,
which will also be described in detail in Section \ref{sec:preliminaries}.

Our goal is to implement access control using IOTA to overcome the limitations in DCACI and to achieve higher flexibility,
finer granularity and higher scalability.
Although the basic idea of using tokens to authorize subjects is similar to DCACI,
we introduce an encryption scheme called the Ciphertext-Policy Attribute-Based Encryption (CP-ABE) \cite{cpabe}
in order to enable more flexible access right authorization.
More specifically, we encrypt the tokens using CP-ABE and distribute them to the subjects through the IOTA Tangle.
More flexible and fine-grained access control can be enforced by finely specifying the policies in CP-ABE.
This greatly facilitates token management of object owners and improves the scalability.

A preliminary version of the proposed scheme has been published as a conference paper in \cite{nakanishi2020iota}.
In comparison with our previous work and DCACI, the main contributions of this paper are as follows:

\begin{itemize}
\item
    Compared with DCACI, the proposed scheme provides stronger security,
	such as secure communication between the subject and object owner when sending access requests as well as
	the authentication of subjects when using their tokens.
	The concrete methods to implement such security measures are also shown,
	whereas in \cite{dcaci} the security measures are not provided.
\item
    Although the idea to use tokens to authorize the subjects is the same as in DCACI, the proposed scheme provides easier token management by
	introducing one-to-many access control, i.e., one token can be used for multiple subjects, whereas in DCACI tokens have to be issued for every subject.
	This greatly facilitates token management of object owners and improves the scalability. In addition, the introduction of CP-ABE enables attribute-based access 	control, which is more fine-grained than that achieved by the DCACI.
\item
	Compared with our previous work, this paper includes enhanced security by introducing an additional security measure,
	i.e., authentication of the subjects in order to prevent the use of illegally-obtained tokens.
	In addition, a more practical prototype of the proposed scheme is implemented with actual IoT devices.
	Moreover, detailed evaluation and analysis of the proposed scheme, such as the relationship between the granularity of access control and
	execution time are carried out, both of which have not been discussed in our previous work.
\end{itemize}

The remainder of this paper is structured as follows. We first introduce some related work in Section \ref{sec:relatedwork}.
We then describe the preliminaries including the DCACI scheme in Section \ref{sec:preliminaries}.
We illustrate our scheme in Section \ref{sec:proposed_scheme} and show its implementation in Section \ref{sec:implementation}.
We then evaluate the performance of our scheme in Section \ref{sec:performance_evaluation}.
Finally, we summarize the paper in Section \ref{sec:conclusions}.

\section{Related Work}\label{sec:relatedwork}
\subsection{Conventional Access Control Schemes}
Various access control models to decide the access rights of subjects and access control policies have been proposed,
among which Access Control List (ACL), Role-Based Access Control (RBAC),
Attribute-Based Access Control (ABAC) and Capability-Based Access Control (CapBAC) are representative ones.
ACL is a table associated with an object that describes what subject can perform what kind of actions on the object \cite{sandhu1994access}.
In RBAC, subjects are first assigned roles (e.g., guest and administrator), and the access rights are determined for each role \cite{sandhu1996role}.
In ABAC, a set of rules, called policies, are defined using the subjects' attributes (e.g., age and affiliation)
and the objects' attributes (e.g., identifier and location) \cite{hu2015attribute}.
In CapBAC, some kind of tokens (e.g., keys and tickets) are issued to subjects as proof of authority (i.e., capabilities) \cite{gusmeroli2013capability}.
The subjects present their tokens to the object owners when accessing the objects.

Based on these models, a variety of conventional access control schemes have been proposed \cite{bhatt2017access, gusmeroli2012iot, liu2012authentication},
most of which fail to cope with large-scale access control for two main reasons.
First, the access rights and policies are stored and managed on a centralized server, which turns out to be a single point of failure.
This makes the system vulnerable to disorder caused by disasters and attacks by malicious users \cite{ouaddah2017access, weber2010internet}.
In the case of attacks, the tampered access rights can lead to an illegal, unintended access control, e.g., unauthorized subjects gain access to some resource.
Second, the central server can be a performance bottleneck, failing to process the increasing amount of access requests in large-scale systems.
Therefore, access control schemes must be distributed, reliable and scalable to cope with the explosively-growing IoT era.

\subsection{Blockchain and Smart Contract}
Blockchain is the underlying technology of cryptocurrencies such as Bitcoin \cite{btc} and Ethereum \cite{eth}.
It is a distributed and tamper-resistant database that is managed over P2P networks.
In blockchain, pieces of data representing remittance between peers, called transactions, are collected and encapsulated as blocks
and appended to the blockchain at regular intervals.
One important property of blockchains is their tamper resistance, which is supported by two main features.
One is that every block includes the hash value of the previous block, which makes the blockchain tamper evident \cite{nist2018blockchain}.
When a peer tampers with the content of a block, the hash value of this block changes,
which leads to inconsistency in the hash values of all succeeding blocks.
This means that all succeeding blocks have to be tampered with as well, in order to maintain consistency.
However, recording new data to the blockchain, i.e., adding a block to the blockchain, requires heavy computation and a certain amount of time,
which is the second feature \cite{pilkington2016blockchain}.
The process of adding a new block to the blockchain is called mining, which is driven by a mathematical puzzle called Proof-of-Work (PoW).
Given a block, PoW is the process of finding a random value called nonce, such that
the hash value of the block resulting from the nonce can meet a pre-defined difficulty requirement, e.g., starting with a certain number of zeros.
It has been known that there is no efficient way to solve PoW, although it is easy to verify.
Another fascinating property of blockchains is that the data are distributed.
Since every peer stores and updates a copy of the blockchain, the data will not be lost even if some peers stop operating.
Because of these attractive properties, blockchains are suitable for storing access rights and policies in terms of access control.

In addition to storing data, recent blockchain systems including Ethereum can handle computation, which is powered by a functionality called
smart contract \cite{ethsmartcontract}.
The core idea is to store executable codes on the blockchain and make peers execute them in a decentralized manner.
The peers reach consensus on the execution result of the code and also store the results in the blockchain.
This enables distributed and reliable computing, which is suitable for processing access requests in terms of access control.

\subsection{Blockchain-Based Access Control Schemes}
In \cite{xu2018blendcac, xu2019exploration, nakamura2019capbac, nakamura2020exploiting},
Ethereum-based distributed CapBAC schemes were proposed,
whose idea is to manage the subjects' tokens on the Ethereum blockchain using smart contracts.
Thanks to the Ethereum blockchain, the tokens can be stored in a distributed and tamper-resistant manner.
When the object receives an access request from the subject, it invokes the smart contract responsible for verifying the subject's token.
Thanks to the smart contracts, processing the access request, i.e., the decision making of permitting or denying the request,
can be performed in a decentralized and reliable manner.

In \cite{dukkipati2018decentralized, maesa2019blockchain, yutaka2019abac, zhang2020attribute},
blockchain-based distributed ABAC schemes were proposed,
where the attributes of the subjects and objects, as well as the access control policies are managed on the Ethereum blockchain using smart contracts.
Thanks to the Ethereum blockchain, the attributes and policies can be stored in a distributed and tamper-resistant manner.
When the subject makes an access request to some object, whether or not the subject's and object's attributes satisfy the corresponding policy
is verified by using smart contracts, enabling decentralized and trustworthy access control.

Ethereum-based RBAC schemes were also developed in \cite{cruz2018rbac, rahman2020context}, whose idea is to manage the association between subjects and roles and also the association between roles and access permissions by smart contracts. In this way, resource owners can decide which subjects can access the resource by referring to the stored associations. Access control schemes based on other models (e.g., ACL) or other blockchain platforms (e.g., Bitcoin) can be found in
\cite{zhang2019IoT, sultana2020data, novo2018blockchain, maesa2017blockchain, ouaddah2016fairaccess, pinno2017controlchain, ding2019novel, zhu2018tbac}.
For detailed introduction, please refer to these references.

As described above, blockchain technology along with smart contracts can solve the limitations in conventional access control schemes and
realize trustworthy and distributed access control. However, the underlying blockchain technology has also introduced new challenges,
i.e., low throughput and high transaction fee. These two problems can impose a great burden on both the administrators and users
when enforcing access control in large-scale environments with a large number of users and highly frequent access requests.

\section{Preliminaries}\label{sec:preliminaries}
\subsection{IOTA}
\begin{figure}[!t]
\centering
\includegraphics[width=8cm]{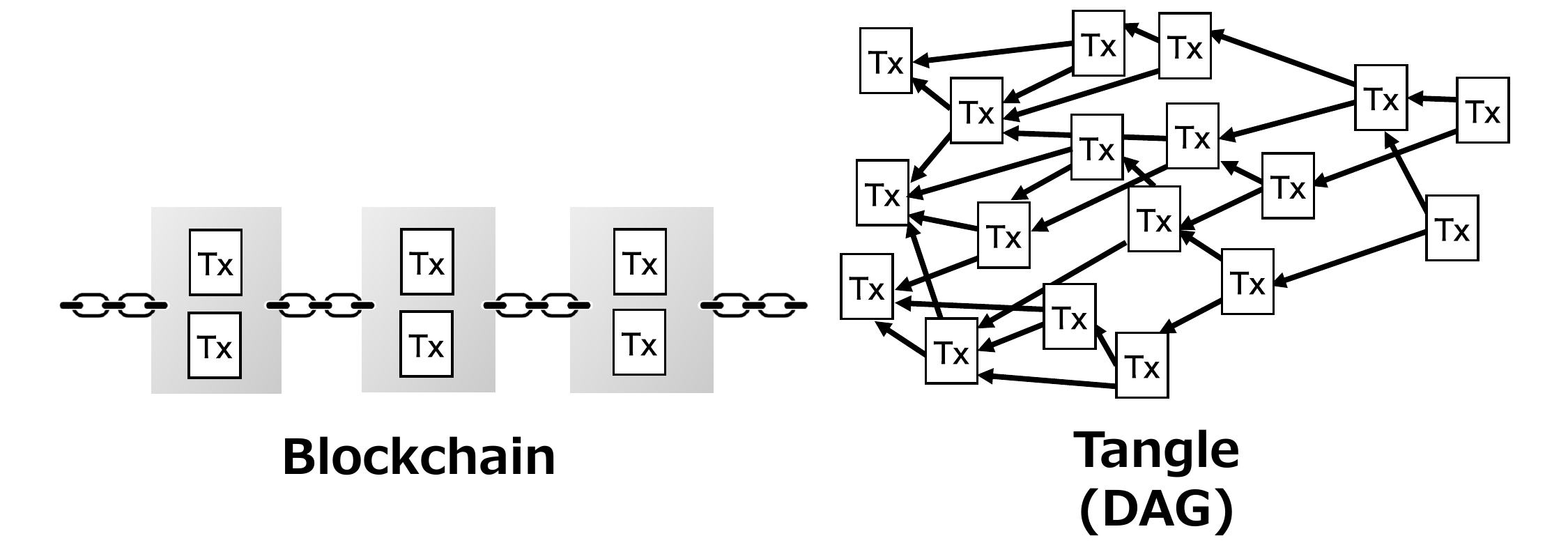}
\caption{Comparison between blockchain and Tangle.}
\label{fig:tangle}
\end{figure}

IOTA is a next-generation distributed ledger technology designed for the IoT.
Unlike blockchain-based cryptocurrencies like Bitcoin and Ethereum, the distributed ledger of IOTA, called the Tangle,
forms a Directed Acyclic Graph (DAG) as shown in Fig.~\ref{fig:tangle}.
Transactions (Txs) are linked together directly using one-way hash functions, instead of being encapsulated in blocks.
Since the limit of block size is removed, newly incoming data can be recorded with high throughput.
Another significant characteristic of IOTA is the removal of mining.
In IOTA, the consistency of the Tangle is maintained by requiring every peer to verify and approve two existing transactions to issue a new transaction.
This not only eliminates the need of transaction fees but also accelerates the speed at which new transactions are approved,
because the increase of the number of incoming transactions leads to more existing transactions being approved.
For these reasons, IOTA is considered to have the potential to solve the problems of high transaction fee and low throughput of blockchain.

\subsection{Masked Authenticated Messaging (MAM)}
\begin{figure}[!t]
\centering
\includegraphics[width=9cm]{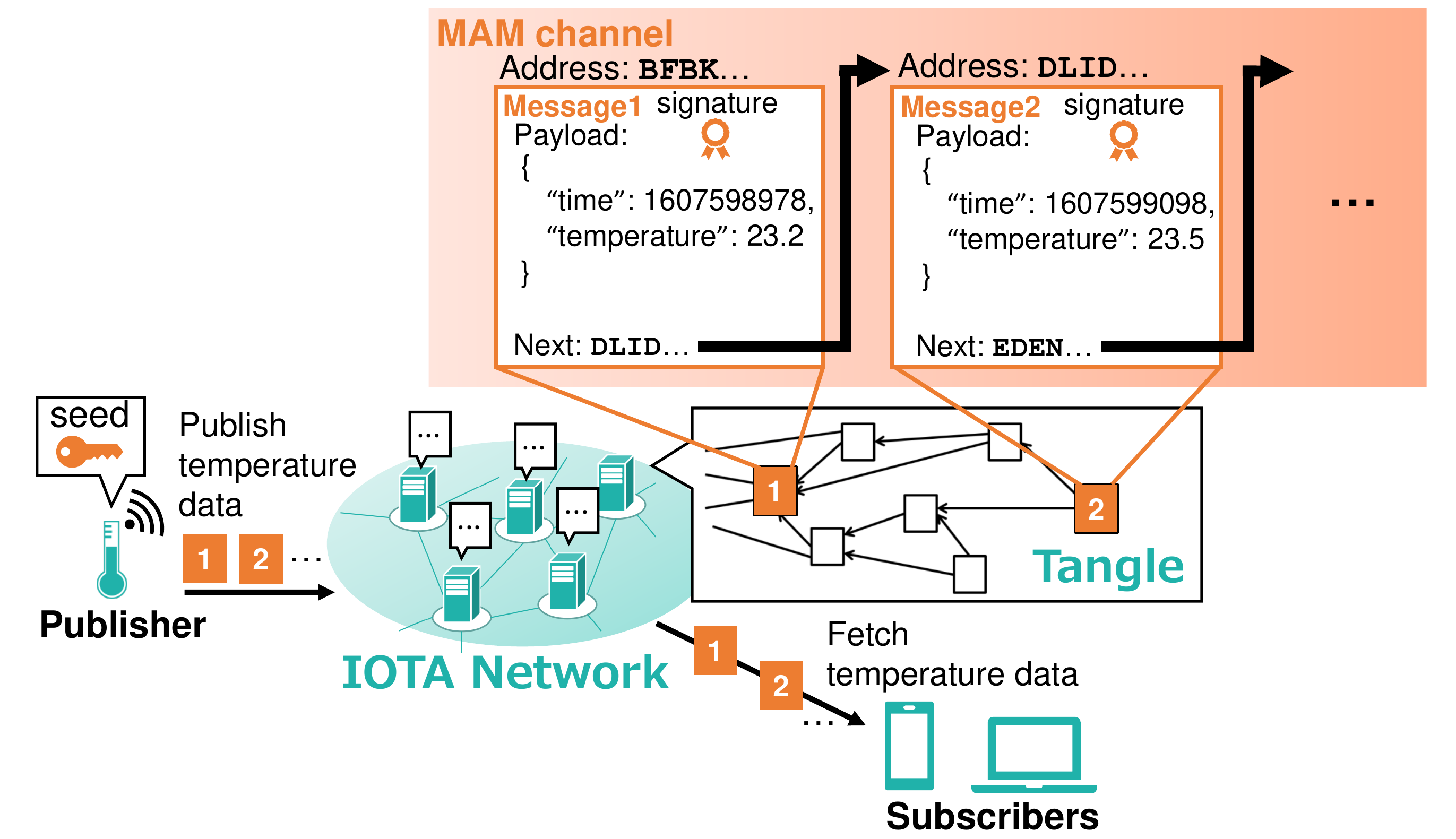}
\caption{MAM channels.}
\label{fig:mam_channels}
\end{figure}

Although smart contracts have not been implemented in IOTA, a novel data communication protocol called Masked Authenticated Messaging (MAM) \cite{mam} is available, which is a promising solution for access control.
Using MAM, peers can record data to and retrieve data from the Tangle in a tamper-resistant fashion.
Peers can record data to the Tangle by masking (i.e., encrypting) them and issuing them as special transactions (MAM transactions).
Each MAM transaction is associated with an address with which any peer can refer to the transaction.
MAM transactions issued by the same peer are linked together chronologically, forming a channel (i.e., a chain of transactions).
MAM channels are useful to record and retrieve sequential data, like periodic recording of temperature data from a smart sensor device,
as illustrated in Fig.~\ref{fig:mam_channels}.
In addition, a signature of the issuer is attached to every MAM transaction, which enables subscribers to verify the authenticity of the issuer (authenticated messaging).
A master password that every peer in the IOTA network keeps, called seed, is used to generate the
addresses and signatures. Only the owner of the seed can publish messages to his/her channel.
With MAM, peers can safely exchange data via the Tangle by subscribing to each other's channel.
This enables us to use the Tangle as a database for storing access rights, in terms of access control.

\subsection{DCACI Scheme}\label{sec:dcaci_scheme}
In this subsection we introduce the DCACI scheme in details. The scheme provides four main operations, i.e.,
\emph{GrantAccess}, \emph{UpdateAccess}, \emph{DelegateAccess} and \emph{GetAccess},
whose functions are authorizing, updating, delegating and verifying access right, respectively.
Since our scheme does not support access right delegation, we focus on the others in this paper.

\subsubsection{\emph{GrantAccess}}
During the initial authorization process, the subject first sends a request for access rights to the object owner.
After authenticating the subject, the owner decides the access rights to grant based on local authorization policies
and issues the subject an access token.
At the same time, the owner records the token to the Tangle as the original copy using MAM.
A new MAM channel is generated for every subject to enable access right update as described below.

\subsubsection{\emph{UpdateAccess}}
When there is a change in the local authorization policies, the owner can update the tokens recorded on the Tangle.
The owner issues a new token and attaches it to the corresponding MAM channel as the next message.
Therefore, the last message in the MAM channel reflects the latest access rights of the corresponding subject,
and the owner will always refer to the last message when validating the subject's access rights.

\subsubsection{\emph{GetAccess}}
To access an object, the subject sends the owner an access request along with the token.
After authenticating the subject, the owner fetches the original copy of the presented token from the Tangle, i.e.,
the last message in the MAM channel associated with the subject.
Based on the original copy, the owner decides whether or not to grant the requested access.

\subsubsection{Limitations in DCACI}
Although DCACI has achieved fee-less distributed access control thanks to IOTA, the framework still suffers from three drawbacks.
First, it is assumed that secure communication links are established between the subjects and object owners,
while no methods for establishing the links are provided.
Thus, requests and tokens are sent without being encrypted, facing the risk of being leaked to malicious users
when an insecure channel is used.
Second, it supports only one-to-one access control, which means that one token must be recorded for each subject,
i.e., one token per subject, increasing the burden of token management for large-scale IoT systems.
Finally, it provides no concrete implementation of the authorization process on the owner sides,
i.e., the way/model used to decide what access rights should be granted to the subjects before issuing tokens,
and the way to authenticate subjects.
To overcome these drawbacks, we propose a novel access control framework based on IOTA and the CP-ABE technology
to realize more flexible and scalable access control.

\subsection{Ciphertext-Policy Attribute-Based Encryption (CP-ABE)}
CP-ABE \cite{cpabe} is a type of public-key cryptosystems.
Unlike common public-key cryptosystems in which each user possesses a pair of public and private keys,
there is only one public key (the master public key) in CP-ABE and private keys are associated with a set of attributes.
The master public key and private keys are issued to users by an attribute authority,
which associates each private key with the attributes of the corresponding user.
For example, the authority may issue a student in the division of Information Science (IS) a private key corresponding to
the set of attributes \{Division: IS, Role: Student\}, and a staff a private key
corresponding to the set of attributes \{Division: IS, Role: Staff\}.

\begin{figure}[!t]
\centering
\includegraphics[width=9cm]{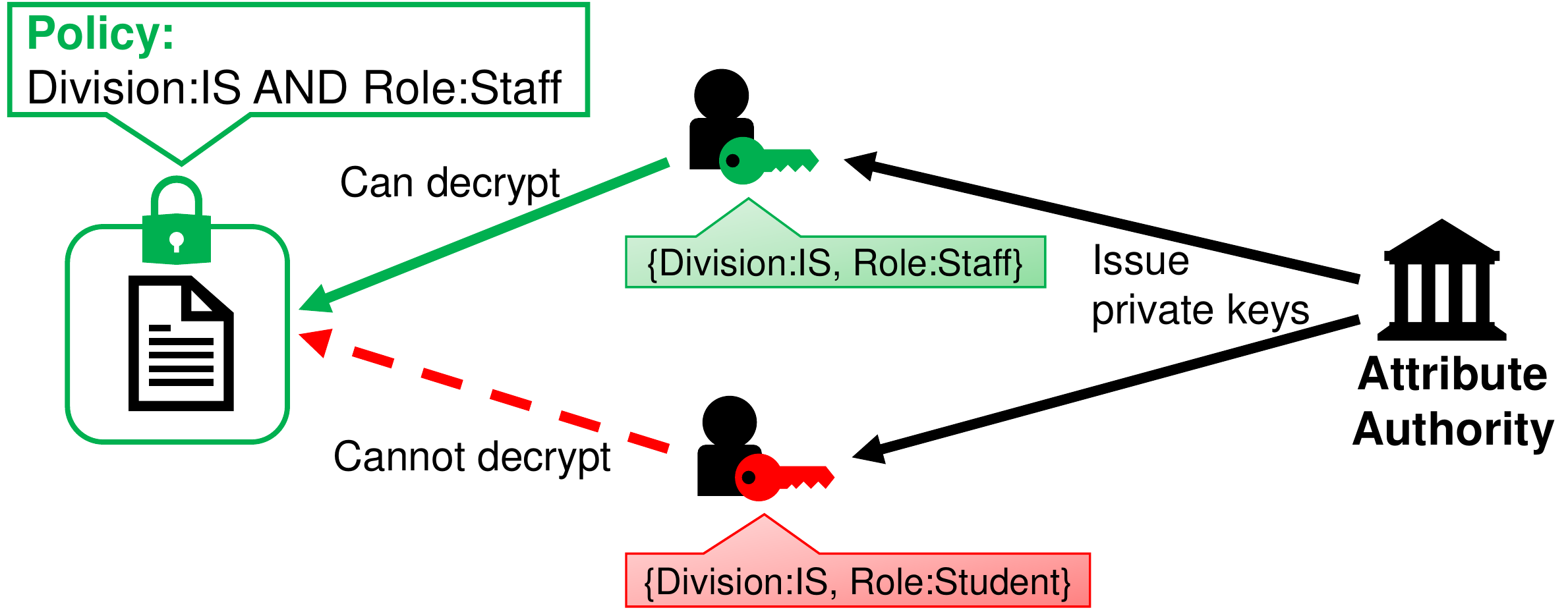}
\caption{Example of access control to data using CP-ABE.}
\label{fig:cpabe}
\end{figure}

Another difference from common cryptosystems is the introduction of logic formulas called policies into the encryption process.
Policies state the conditions of attributes that need to be satisfied to decrypt the ciphertext.
The decryption succeeds if and only if the set of attributes associated with the private key satisfies the policy.
Fig.~\ref{fig:cpabe} shows an example. Given a ciphertext encrypted using the policy ``Division: IS AND Role: Staff,''
users with the private key corresponding to the set of attributes \{Division: IS, Role: Staff\} can decrypt it,
while users with the private key corresponding to the set of attributes \{Division: IS, Role: Student\} cannot.

As seen above, CP-ABE enables fine-grained ABAC to data by restricting the successful decryption to a specific group of users using policies.
In our scheme, we encrypt tokens using CP-ABE and store them on the Tangle to achieve flexible access right authorization.

\section{Proposed Scheme}\label{sec:proposed_scheme}
We combine CP-ABE with IOTA to solve the issues mentioned in Section \ref{sec:dcaci_scheme}.
Access rights are managed on the IOTA Tangle in the form of tokens, which are encrypted using CP-ABE and recorded as MAM transactions.
Our scheme is thus a hybrid of CapBAC and ABAC.

\subsection{Token Structure}

\begin{figure}[!t]
\centering
\includegraphics[clip, width=8.8cm]{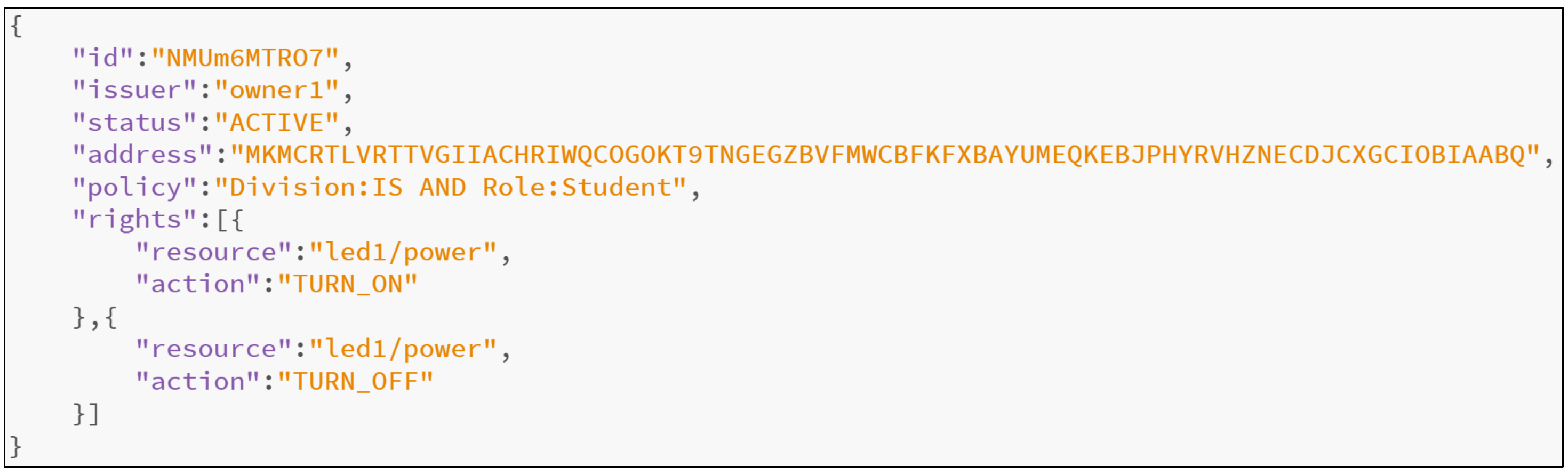}
\caption{Example of a token.}
\label{fig:token}
\end{figure}

As shown in Fig.~\ref{fig:token}, a token is a JSON object issued by object owners.
A token contains
a unique identifier (ID),
the issuer,
the address it is associated with on the Tangle,
the policy that must be satisfied to decrypt the token,
the current status
and a list of access rights.
This indicates that only the subjects whose attributes satisfy the policy in the token
can decrypt the token and are thus granted the access rights recorded in the token.
For instance, the token shown in Fig.~\ref{fig:token}
has been issued by {\tt owner1} and is associated with the address {\tt MKM$\ldots$ABQ} on the Tangle.
Only subjects whose attributes satisfy the policy {\tt Division:IS AND Role:Student}
can perform the two actions {\tt TURN\_ON} and {\tt TURN\_OFF} on the resource {\tt led1/power}.
The status field {\tt ACTIVE} can be changed to {\tt INACTIVE} by the owner afterward in order to revocate access rights, which is
introduced in detail later.
Using this token structure, we design the overall architecture of the proposed scheme.

\subsection{Access Right Authorization}
\begin{figure}[!t]
\centering
\includegraphics[clip, width=9cm]{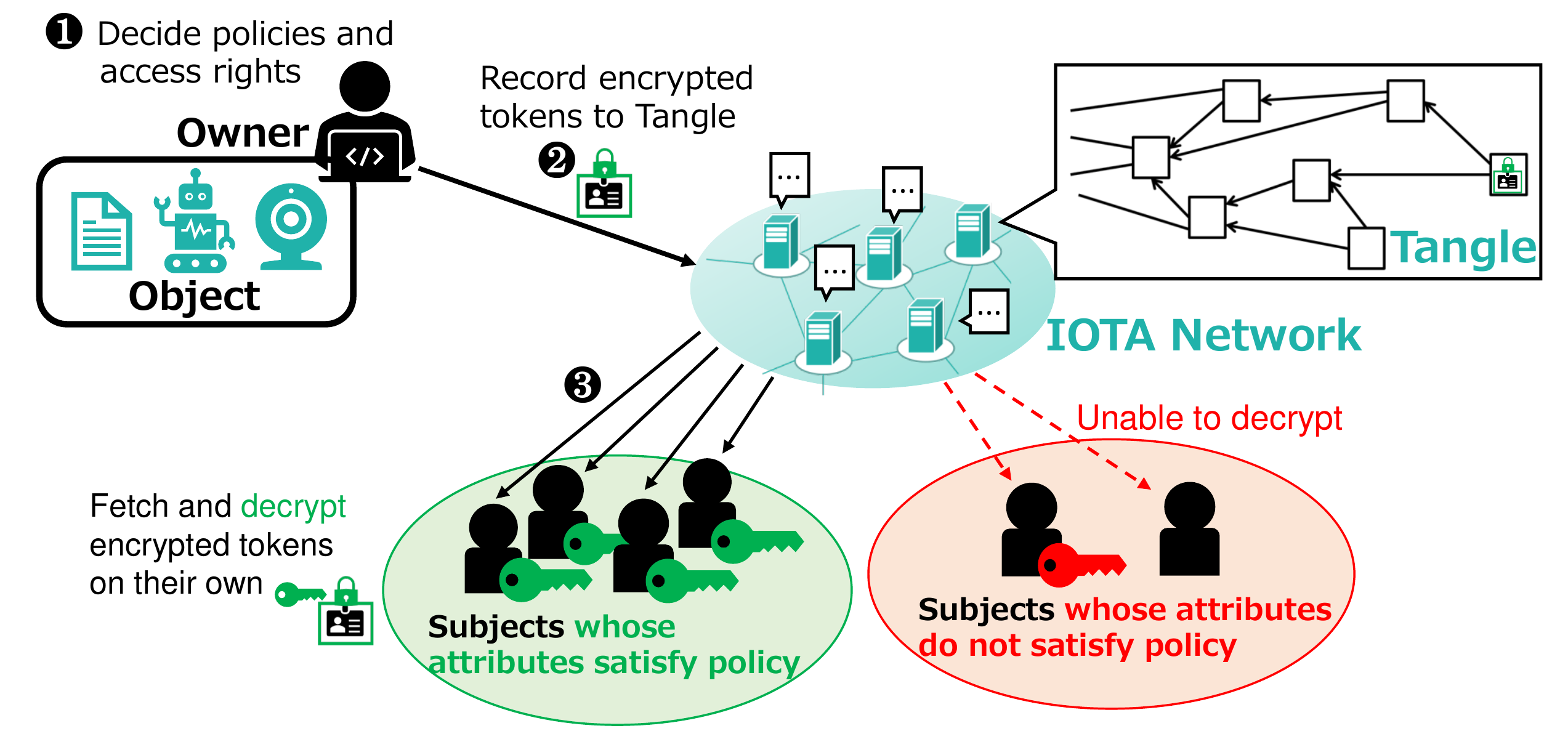}
\caption{Proposed scheme: access right authorization.}
\label{fig:proposed_authorization}
\end{figure}

Fig.~\ref{fig:proposed_authorization} illustrates how access rights are authorized to the subjects.
The object owner first decides the policy to be embedded in the token and the corresponding access rights
to specify which group of subjects can perform what actions to the object (Step 1 in Fig.~\ref{fig:proposed_authorization}).
Examples of policies and access rights are shown in Table~\ref{tab:local_policy}.
The first example implies that subjects satisfying the policy ``Division: IS AND Role: Student'' will be authorized to
perform two actions ``TURN\_ON'' and ``TURN\_OFF'' on the resource ``led1/power.''
After deciding the policy and access rights, the object owner prepares a token according to pre-defined structure
and encrypts it under the policy using CP-ABE.
This means that only those who satisfy the policy can decrypt it.
For the first example in Table~\ref{tab:local_policy}, the object owner will issue a token as shown in Fig.~\ref{fig:token} and then encrypt it under the policy ``Division: IS AND Role: Student.''

\begin{table}[!t]
\caption{Examples of policies and access rights.}
\centering
\includegraphics[clip, width=9cm]{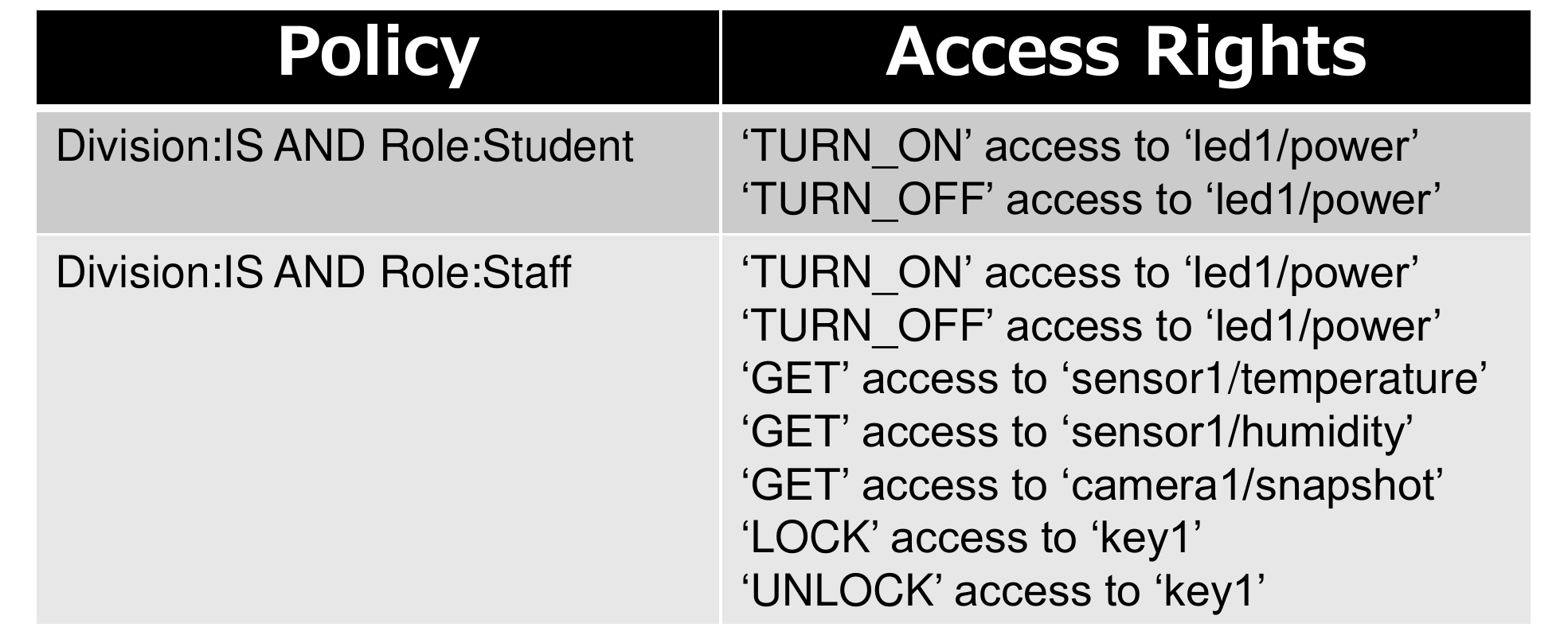}
\label{tab:local_policy}
\end{table}

The owner then records the encrypted token to the Tangle using MAM (Step 2 in Fig.~\ref{fig:proposed_authorization}).
A new MAM channel is generated for every policy so that the token can be updated afterward
when there is any change in the corresponding access rights, as introduced later.
Although the MAM transaction itself is public and visible to any peer,
the token can be decrypted only by those with a private key associated with a set of attributes satisfying the policy.
Subjects can fetch the encrypted token from the Tangle and decrypt it using their private keys
(Step 3 in Fig.~\ref{fig:proposed_authorization}).
It is assumed here that the address associated with the MAM transaction containing the encrypted token is made public and available to the subjects.
In this way, once the owner has recorded an encrypted token to the Tangle,
all subjects satisfying the policy can obtain the token and are authorized the access rights.
On the contrary, in DCACI, the owner has to conduct the authorization (although the implementation is not provided)
and token issuance processes once for every subject, which increases the burden of object owners.
Our scheme not only alleviates the burden of the owner but also enables one-to-many access control.
This means that one token is responsible for the access control of a group of subjects, while, in DCACI, one token is only for one subject.

\subsection{Access Right Update}\label{sec:update_scenario}
\begin{figure}[!t]
\centering
\includegraphics[clip, width=9cm]{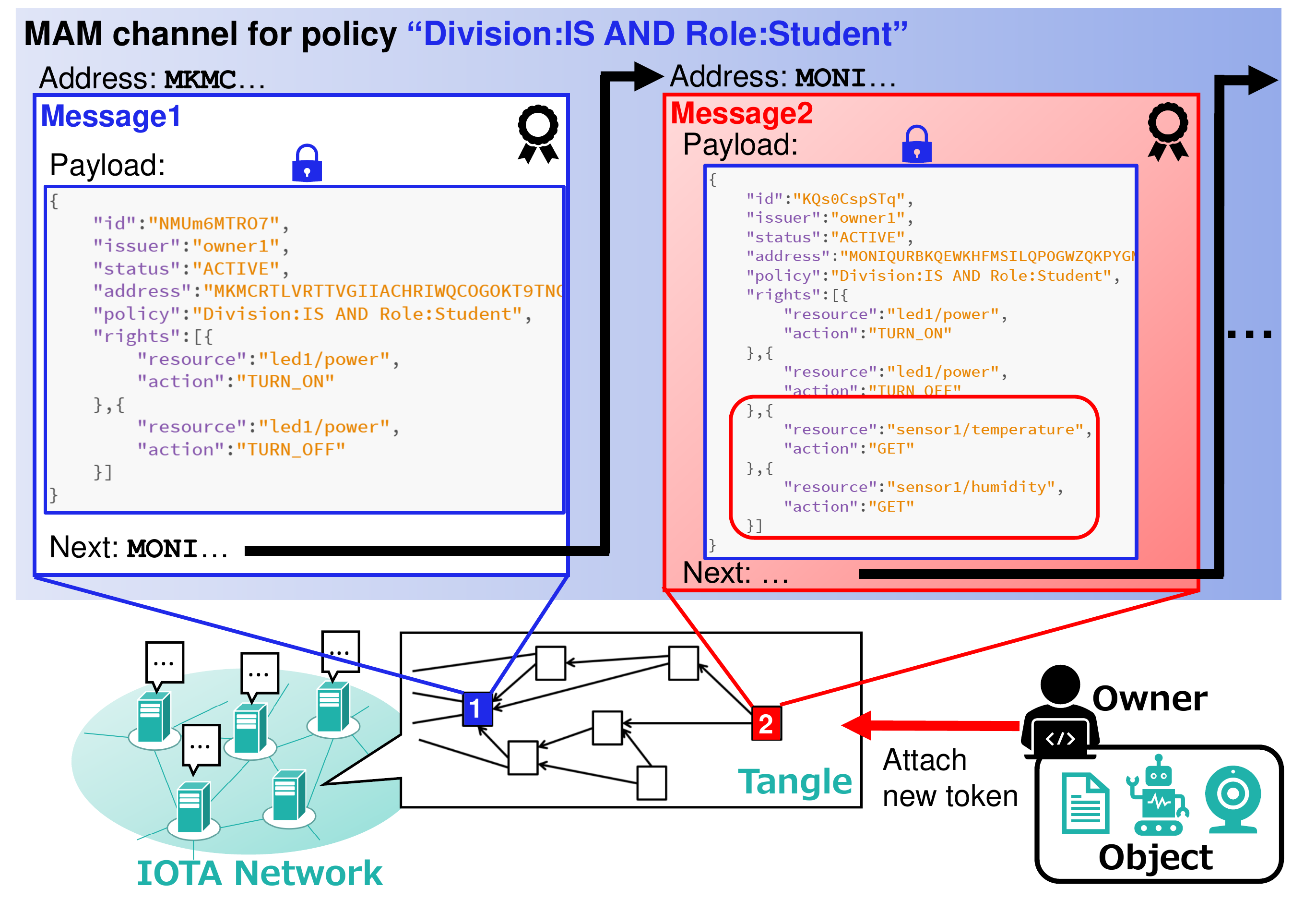}
\caption{Example of access right update.}
\label{fig:proposed_update}
\end{figure}

Fig.~\ref{fig:proposed_update} illustrates how access rights can be updated by the owner.
We use the same approach as in DCACI, i.e., publishing a new token containing the updated access rights as the next message in the corresponding MAM channel.
For example, we consider a case where the owner changes the access rights for policy ``Division: IS AND Role: Student''
from those in Table~\ref{tab:local_policy} to those in Table~\ref{tab:local_policy_updated}.
The owner attaches the new token (encrypted using CP-ABE) to the MAM channel corresponding to policy ``Division: IS AND Role: Student'' as the next message.
Since older tokens no longer hold the latest access rights when updates occur, the owner always refer to the last message in the channel when retrieving access rights from the Tangle.
Likewise, the owner can also revocate tokens by updating the status field of tokens to ``INACTIVE.''

\begin{table}[!t]
\caption{Policies and access rights after update.}
\centering
\includegraphics[clip, width=9cm]{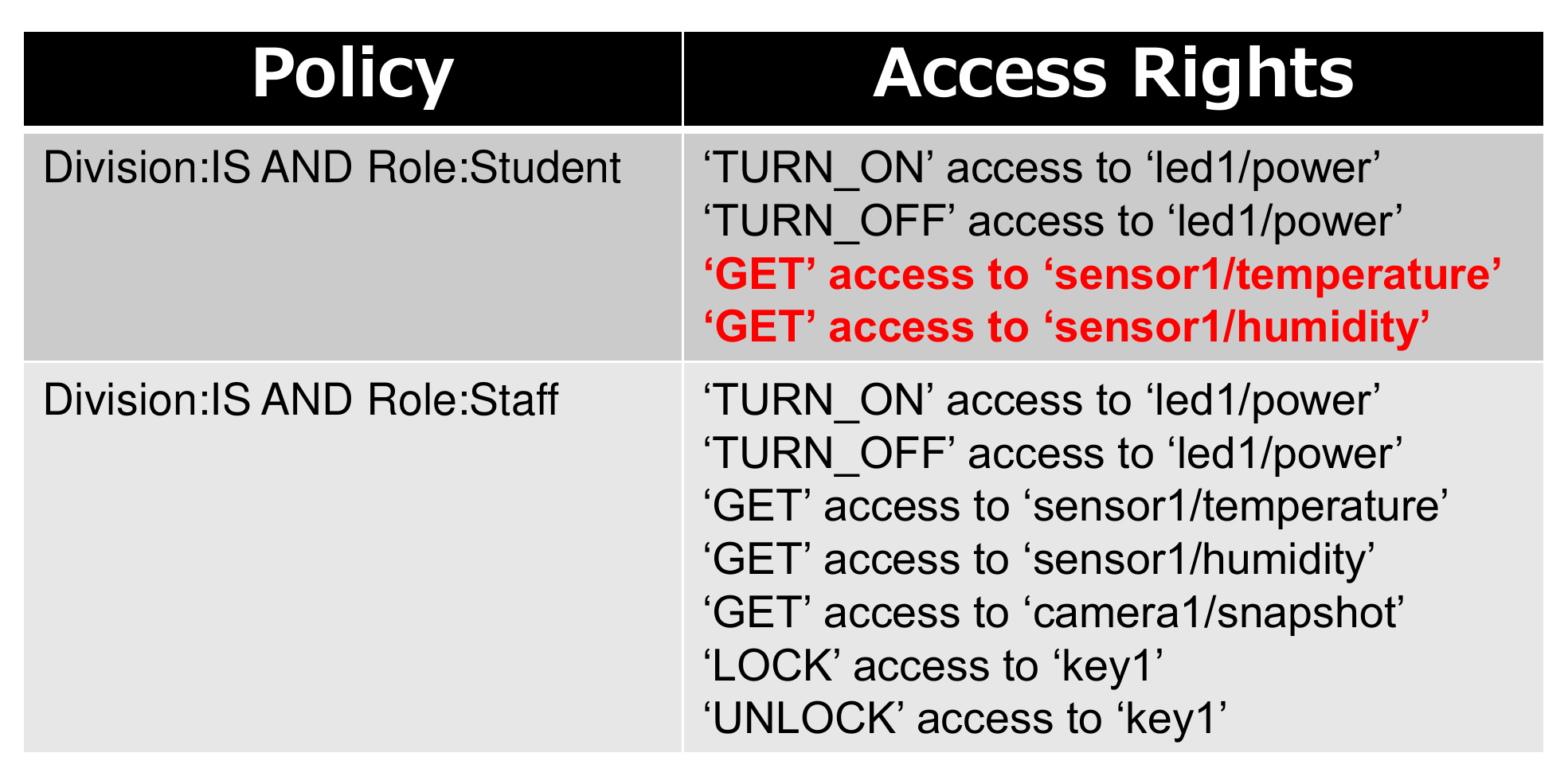}
\label{tab:local_policy_updated}
\end{table}

\subsection{Access Right Verification}
Although the basic idea in our scheme is similar to \emph{GetAccess} in DCACI, i.e., subjects can use their tokens to access resources,
we introduce an original authentication phase in order to prevent the use of illegally-obtained tokens.
This is required because we have to take into account illegal transfer of tokens among subjects and token stolen by malicious subjects.
Although it is mentioned in \cite{dcaci} that such cases can be detected by authenticating the subject since tokens are unique to each subject in the scheme, the concrete method to authenticate subjects is not provided.

The proposed verification process consists of two phases, i.e., the authentication phase and the access request phase.
The former is required to make sure that the subject indeed satisfies the corresponding policy, and the latter is required to verify that the requested action can be executed using the corresponding access rights.

\subsubsection{Authentication Phase}
In the authentication phase, the owner imposes a One-Time Password (OTP) authentication on the subject
in order to verify that the subject, who is about to use some token, satisfies the corresponding policy embedded inside the token.
We design the OTP authentication process such that only subjects with an appropriate private key can obtain the OTP,
and thus those who illegally obtained tokens will not be able to use them.

\begin{figure}[!t]
\centering
\includegraphics[clip, width=9cm]{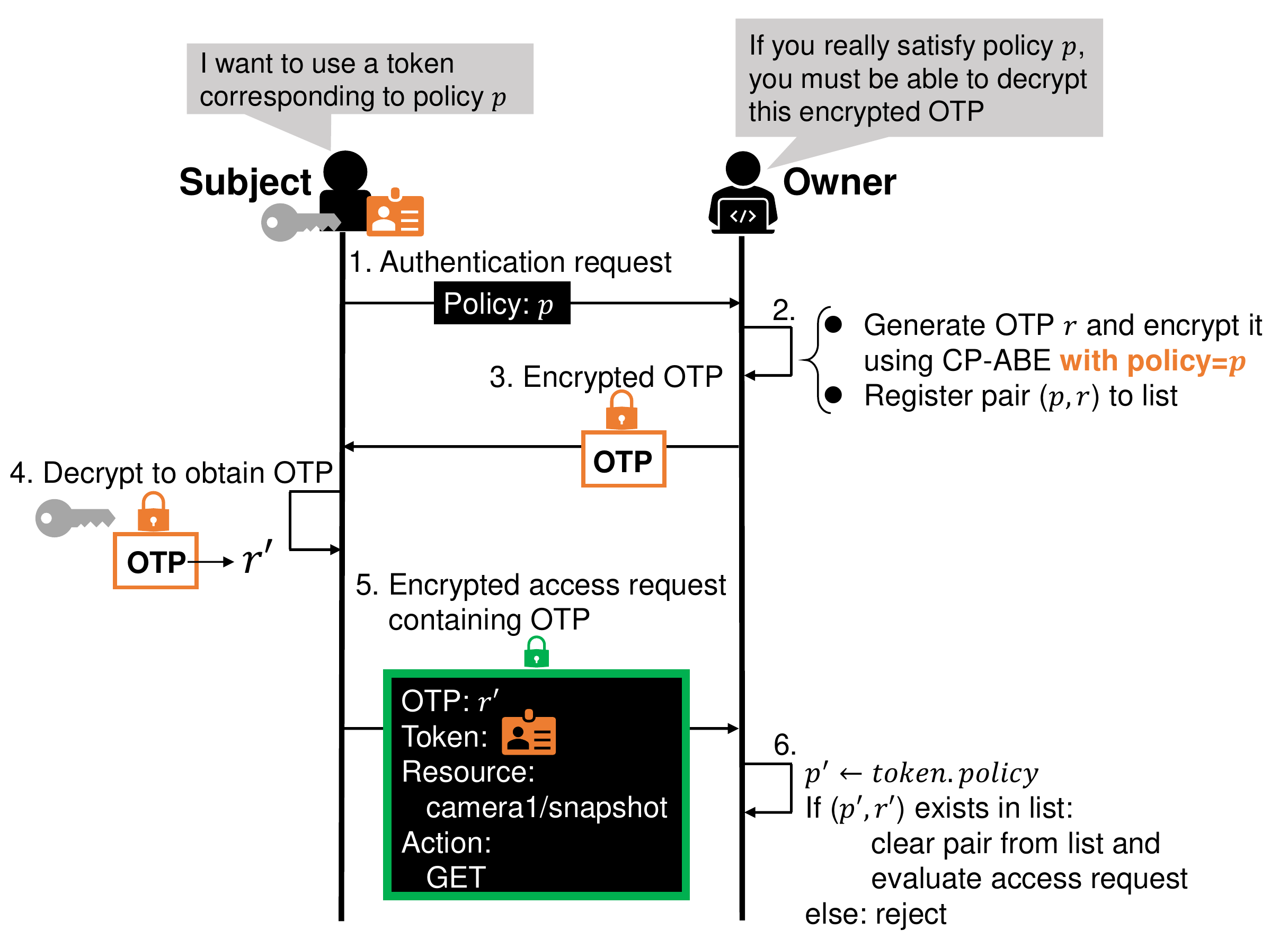}
\caption{Proposed scheme: access right verification (authentication phase).}
\label{fig:proposed_verification_otp}
\end{figure}

The flow of the OTP authentication is depicted in Fig.~\ref{fig:proposed_verification_otp}.
Subjects who want to use a token to access some resource first make an authentication request by declaring the policy corresponding to the token (Step 1 in Fig.~\ref{fig:proposed_verification_otp}).
What the owner wants to verify here is that the subject really satisfies the required policy i.e., holds a private key associated with a set of attributes satisfying the policy.
To do so, the owner sends back an OTP encrypted under the policy using CP-ABE (Steps 2 to 3 in Fig.~\ref{fig:proposed_verification_otp}), which means that only those who satisfy the policy can decrypt it.
At the same time, the owner registers the policy-OTP pair to a list for later confirmation.

On receiving the encrypted OTP, the subject decrypts it using his/her private key
and presents it to the owner through an encrypted access request (Steps 4 to 5 in Fig.~\ref{fig:proposed_verification_otp}),
along with the resource to access, the action to perform and the token to present.
The request is encrypted using CP-ABE, ensuring secure communication between the subject and owner.
The subject encrypts the request under some policy that allows only the object owner to decrypt it (e.g., ``Role: Owner'').

On receiving the encrypted access request, the owner decrypts it using his/her private key issued by the authority.
The owner extracts the OTP from the access request and the policy from the presented token, and then checks the pair against the policy-OTP pair list registered in Step 2
(Step 6 in Fig.~\ref{fig:proposed_verification_otp}).
If the pair exists in the list, the OTP is considered valid and cleared from list, after which the owner proceeds to the request phase.
Otherwise, the OTP is invalid and the request is thus rejected.

In this way, subjects who illegally obtained the token will not be able to obtain the valid OTP and thus cannot generate a valid access request,
meaning that they cannot use the token.

\subsubsection{Access Request Phase}
After the authentication phase, the owner evaluates the access request based on the presented token.
The owner first fetches the original copy of the corresponding token (i.e., the one issued during authorization) from the Tangle.
The presented token is checked against the original token to verify its authenticity.
Since the Tangle is tamper-proof, any modification in the token can be detected in this process.
The access request is rejected if the token verification fails.
If the token is valid, the resource and action are evaluated based on the list of access rights contained in the token.
If the requested action does not exist in the list, the request is rejected because it is an attempt of unauthorized access.

\section{Implementation}\label{sec:implementation}
To show the feasibility of our scheme, we have implemented a proof-of-concept prototype using the IOTA Mainnet \cite{mainnet} and IoT devices.

\subsection{System Configuration}
\begin{figure}[!t]
\centering
\includegraphics[clip, width=8.8cm]{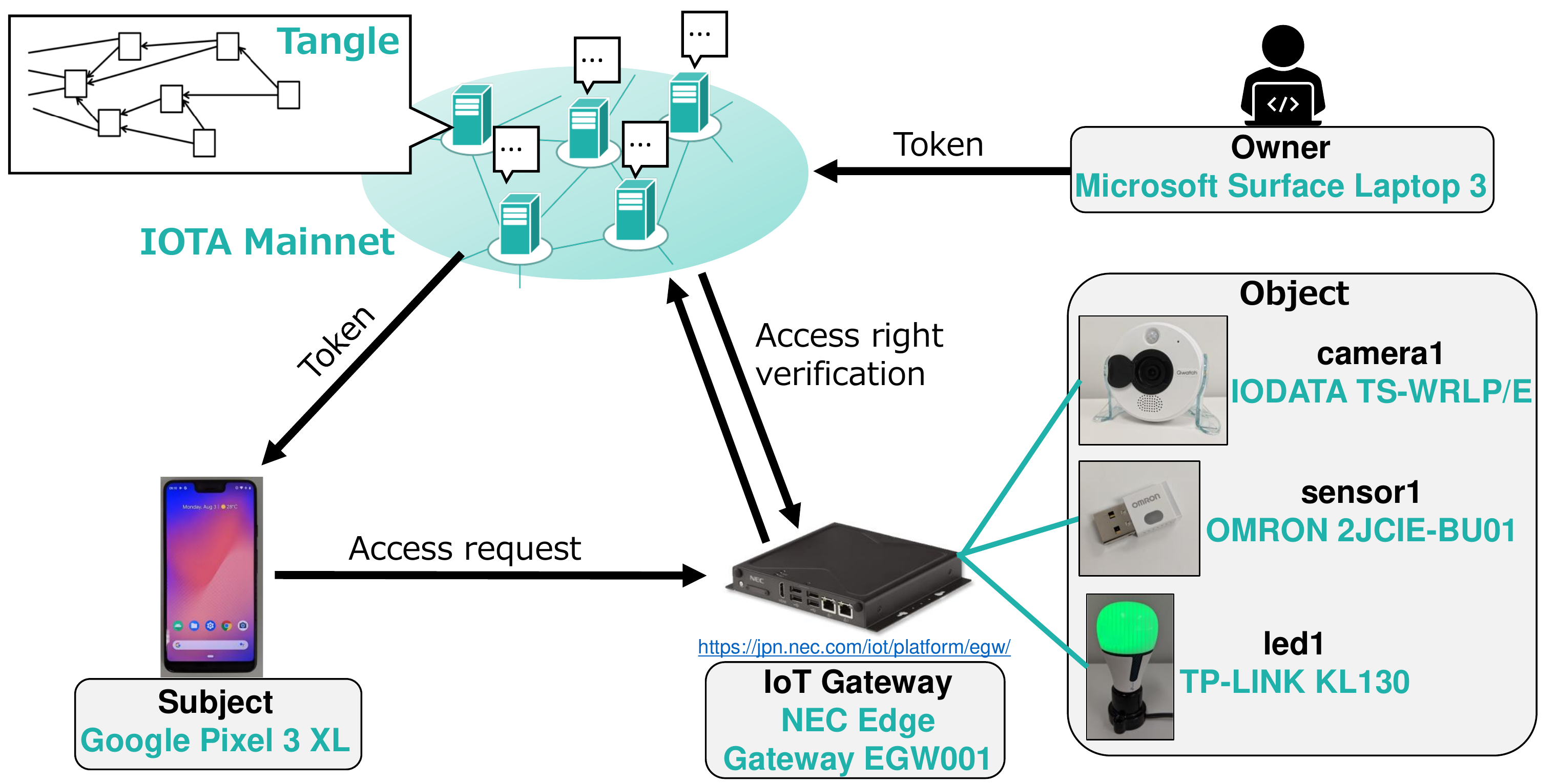}
\caption{Configuration of the prototype system.}
\label{fig:prototype_configuration}
\end{figure}

As shown in Fig.~\ref{fig:prototype_configuration},
we used a Microsoft Surface Laptop 3 (1.5 GHz Intel Core i7, 16GB RAM) as the object owner
and a Google Pixel 3 XL (2.5 GHz Qualcomm Snapdragon 845, 4GB RAM) as the subject.
The owner manages three objects, ``camera1'' (IODATA TS-WRLP/E), ``sensor1'' (OMRON 2JCIE-BU01) and ``led1'' (TP-LINK KL130).
The objects can be accessed through the IoT gateway (NEC EGW001: 1.46 GHz Intel Atom, 2GB RAM) which performs access right verification.
The official JavaScript API \cite{mamapi} was used to issue and fetch MAM transactions and
implementation based on \cite{zlwen} was used to handle the CP-ABE encryption and decryption.
The gateway service was implemented by an HTTP server which listens to requests from the subject.
At the subject side, we developed a native Android application which can
fetch MAM transactions from the Tangle, handle CP-ABE and fire access requests.
All entities participate in the IOTA Mainnet as clients and communicate with a full
node ({\tt https://nodes.thetangle.org}) to issue and fetch MAM transactions.

\subsection{Access Right Authorization}
\begin{figure}[!t]
\centering
\includegraphics[clip, width=8.7cm]{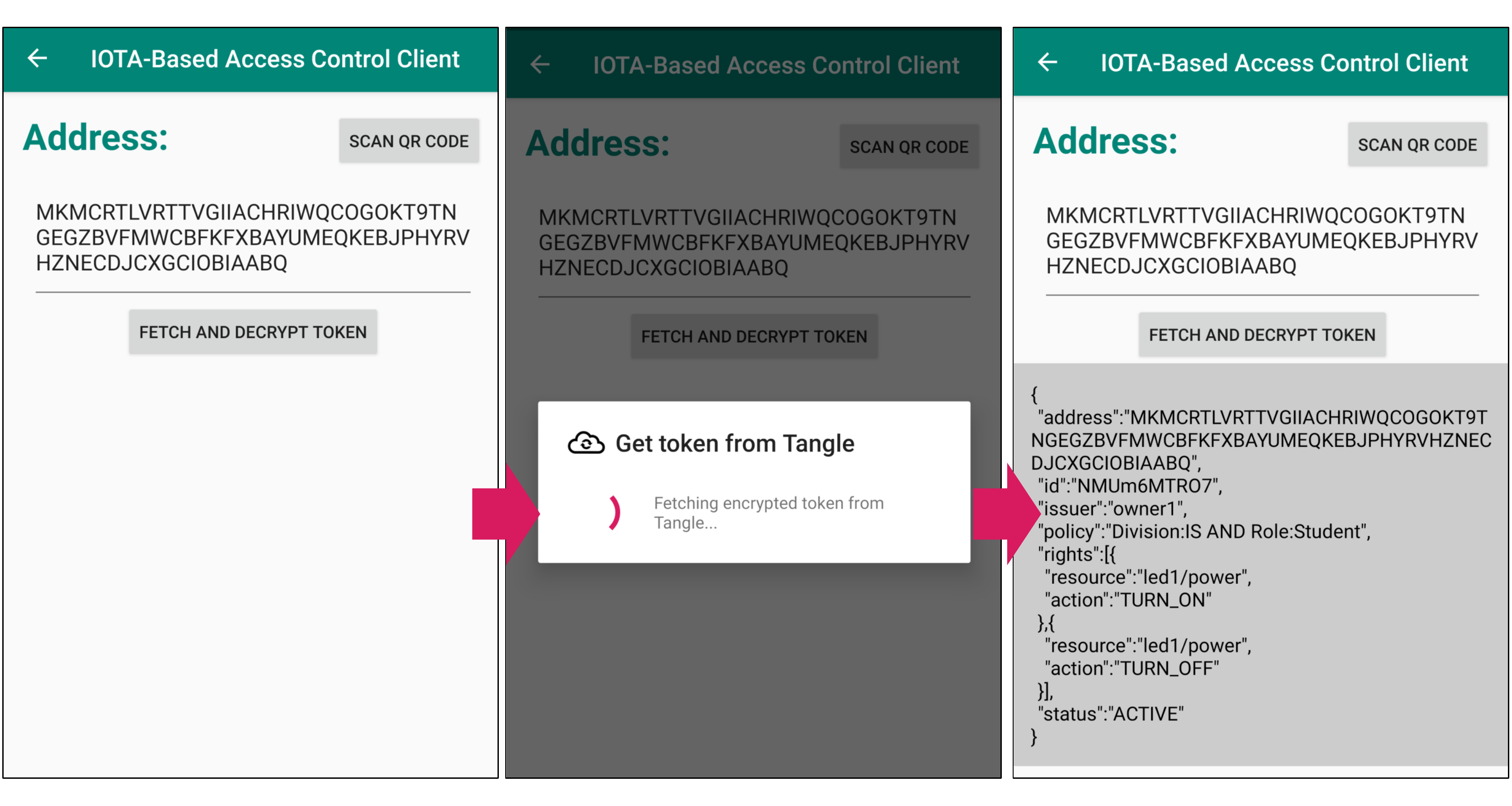}
\caption{Token fetching from Tangle and decryption.}
\label{fig:prototype_authorization}
\end{figure}

For simplicity, we limited the tokens to the two illustrated in Table~\ref{tab:local_policy_updated} (i.e., the student token and the staff token)
and issued the subject a private key associated with the set of attributes \{Division: IS, Role: Student\}.
Fig.~\ref{fig:prototype_authorization} shows the result of fetching and decrypting the student token using the private key.
We can see that the access right is successfully authorized via the Tangle to the subject.
Decrypted tokens will be stored on local storage as text files, which can be used to access resources later.

\subsection{Access Right Update}
\begin{figure}[!t]
\centering
\includegraphics[clip, width=8.7cm]{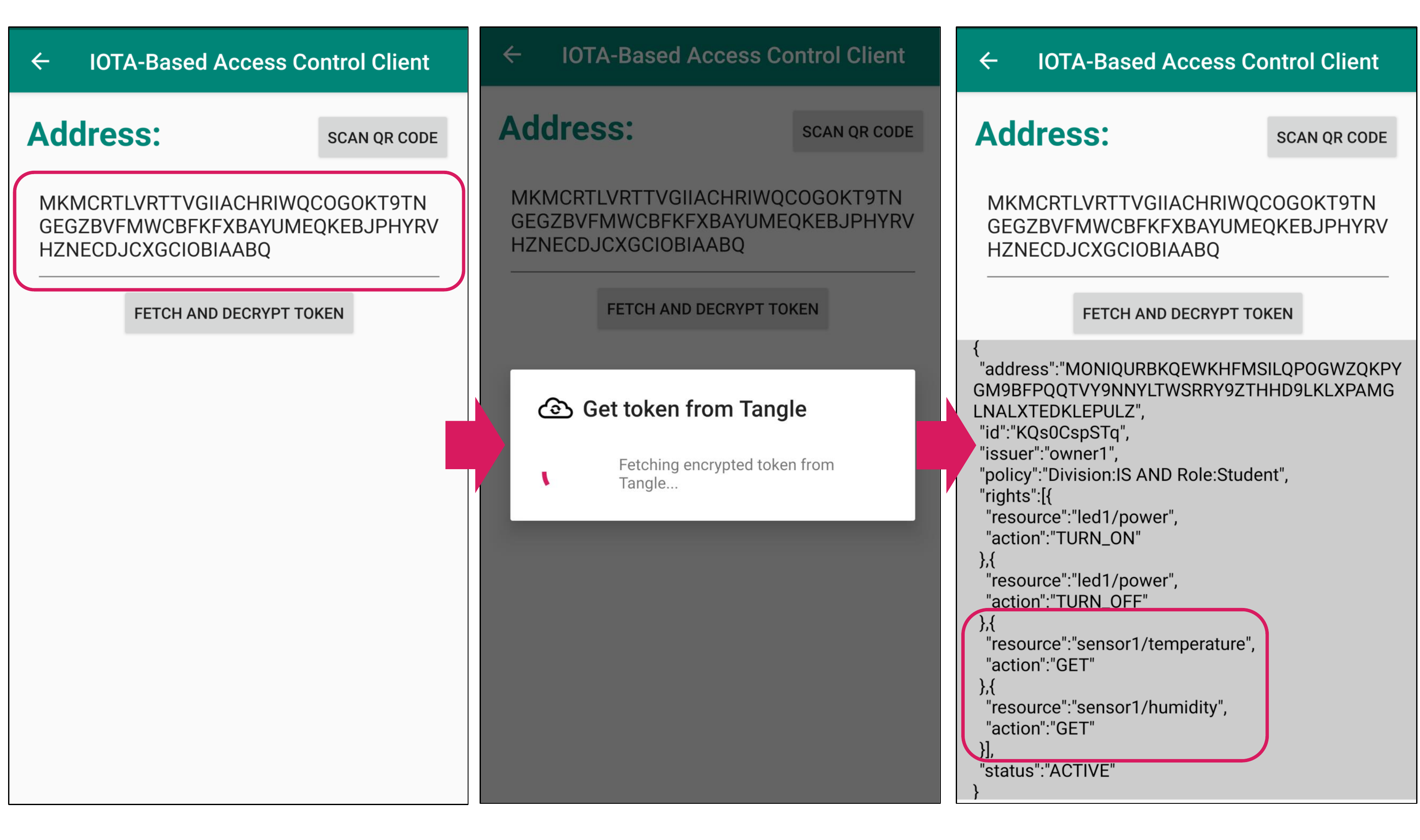}
\caption{Token fetching from Tangle and decryption after access right update.}
\label{fig:prototype_update}
\end{figure}

Fig.~\ref{fig:prototype_update} shows the result of fetching and decrypting the student token
\emph{after} the access right update mentioned in Section \ref{sec:update_scenario}.
The application walks through the MAM channel starting from the first message which contains the first token shown in Fig.~\ref{fig:prototype_authorization}
(note that the address is the same as in Fig.~\ref{fig:prototype_authorization}) and fetches the latest message, i.e., the second message, which contains the updated token.
We can see that the access right is successfully updated (the access right pertaining to ``sensor1'' has been added)
and that the subject can always obtain the latest token via the Tangle.

\subsection{Access Right Verification}
\begin{figure}[!t]
\centering
\includegraphics[clip, width=9cm]{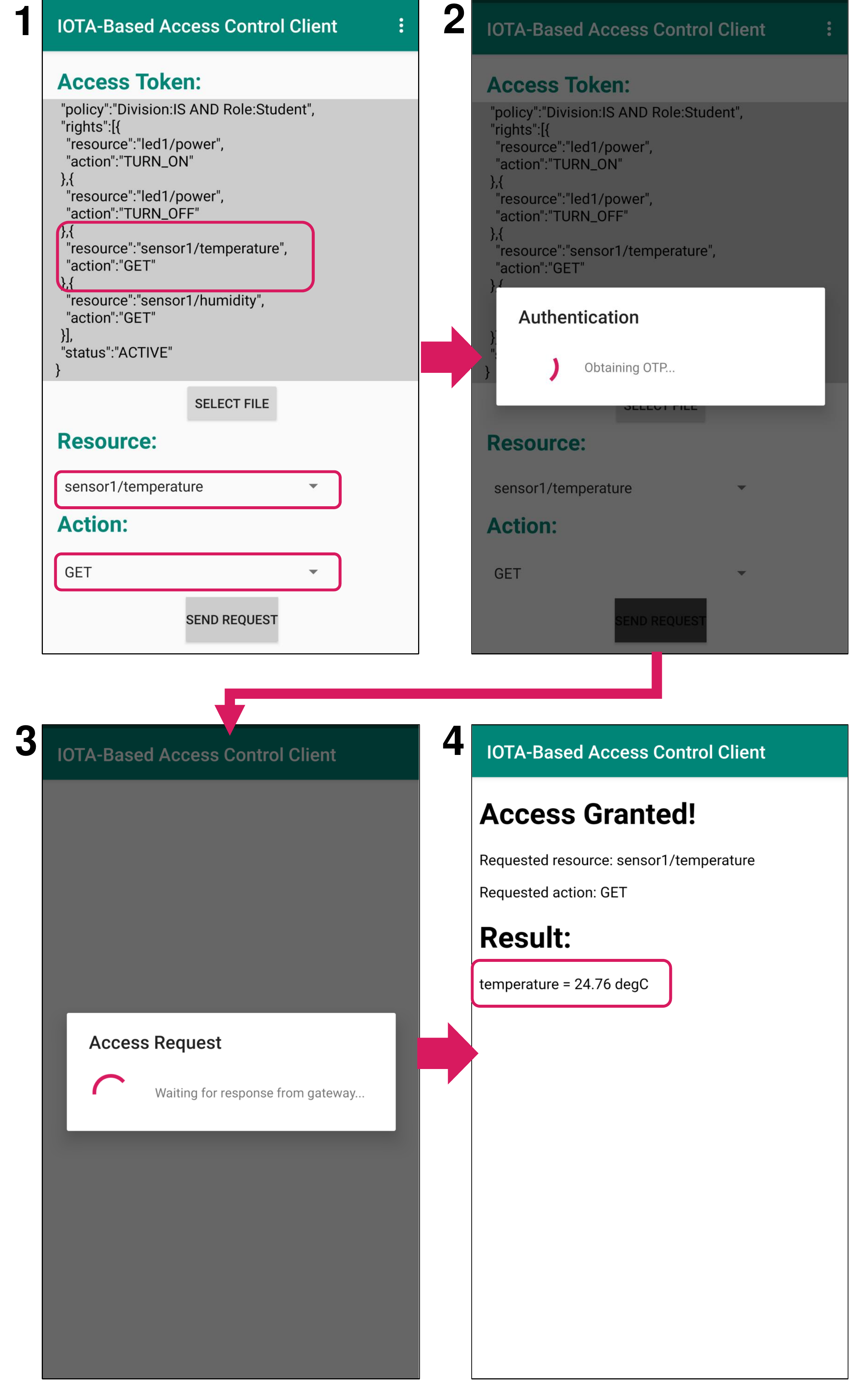}
\caption{Granted access.}
\label{fig:granted_access}
\end{figure}

Fig.~\ref{fig:granted_access} shows the result of a student accessing a resource using the student token and his/her CP-ABE private key.
The subject (i.e., the student) first selects the token to present (the one saved to local storage during authorization), the resource to access (``sensor1/temperature'') and the action to perform (``GET'') (Image 1 in Fig.~\ref{fig:granted_access}).
By pressing the ``SEND REQUEST'' button, the authentication phase is initiated,
where an authentication request with the corresponding policy ``Division:IS AND Role:Student'' is sent to the owner
(Image 2 in Fig.~\ref{fig:granted_access}).
In this case, authentication succeeds since the subject holds a private key associated with the set of attributes \{Division: IS, Role: Student\} as mentioned earlier,
and is thus able to decrypt the encrypted OTP sent back from the owner.
Therefore, the subject can proceed to the access request phase (Image 3 in Fig.~\ref{fig:granted_access}).
In the access request phase, the presented token is authentic and it contains the right to perform the action ``GET'' on the resource ``sensor1/temperature.''
Therefore, access is granted (Image 4 in Fig.~\ref{fig:granted_access}).

\begin{figure}[!t]
\centering
\includegraphics[clip, width=9cm]{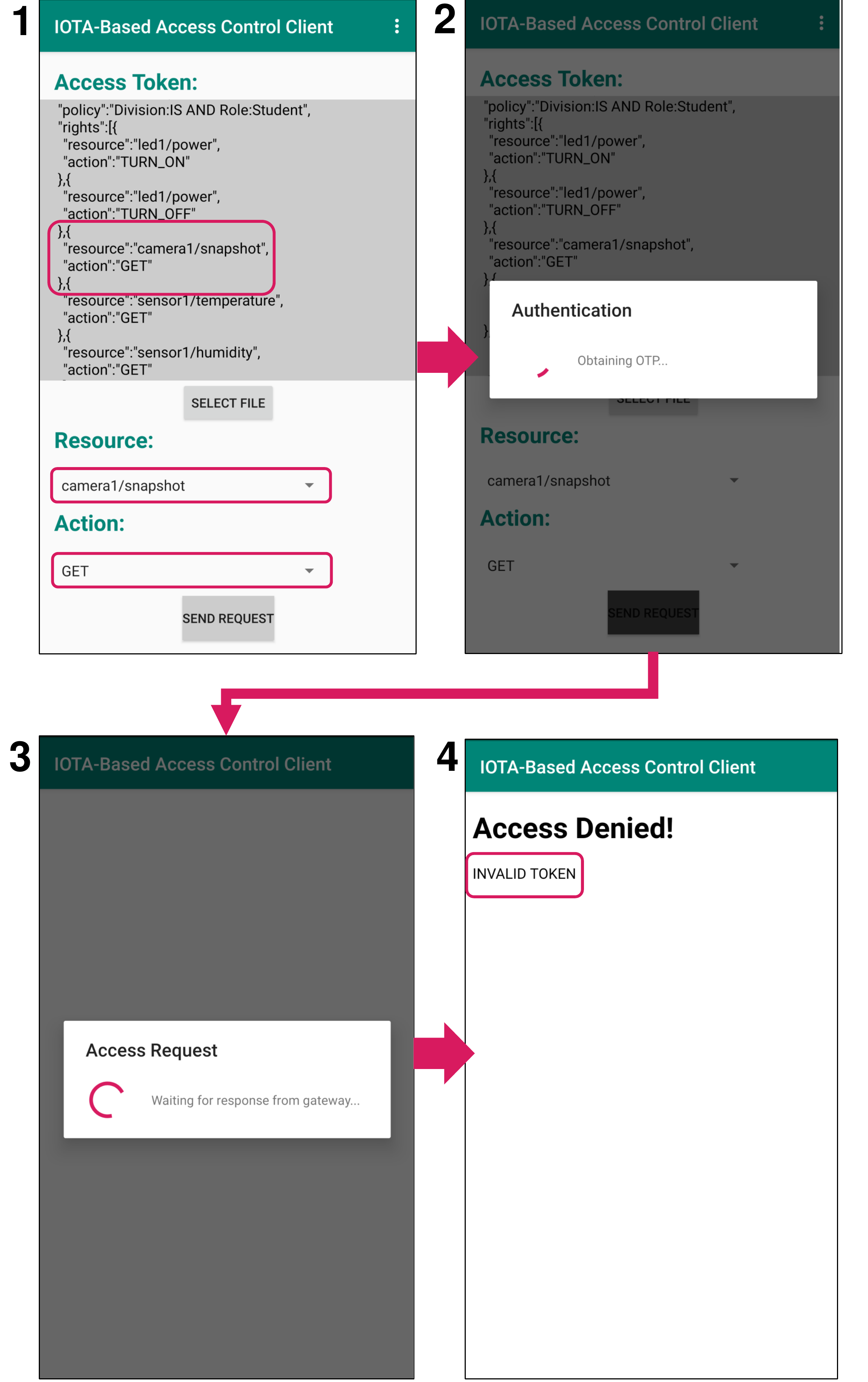}
\caption{Denied access due to tampered token.}
\label{fig:denied_tampered}
\end{figure}

On the other hand, Fig.~\ref{fig:denied_tampered} shows the result of an attempt by a student
to perform the action ``GET'' on the resource ``camera1/snapshot'' (which is exclusive to the staff) using a tampered token.
Based on the student token, a new ``rights'' clause containing ``camera1/snapshot'' and ``GET'' as the resource field and action field, respectively, has been added to the token (Image 1 in Fig.~\ref{fig:denied_tampered}).
In the authentication phase, an authentication request with the corresponding policy ``Division:IS AND Role:Student'' is sent to the owner
(Image 2 in Fig.~\ref{fig:denied_tampered}).
Although the subject can pass the authentication in the same manner as the granted case,
the owner detects the modification on the token in the request phase by checking the presented token against the original token fetched from the Tangle.
Therefore, access is rejected (Image 4 in Fig.~\ref{fig:denied_tampered}).

\begin{figure}[!t]
\centering
\includegraphics[clip, width=9cm]{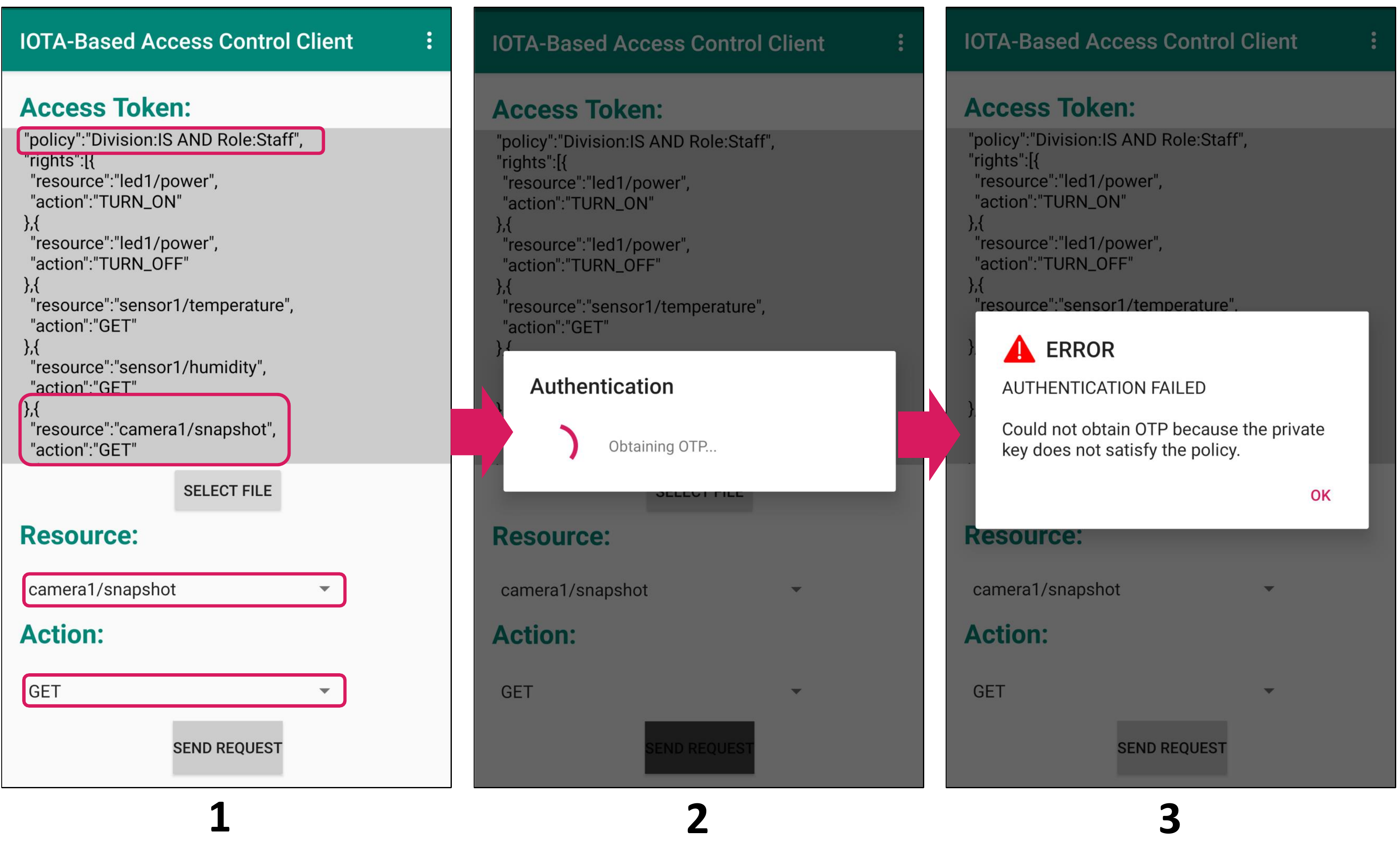}
\caption{Denied access due to stolen token.}
\label{fig:denied_stolen}
\end{figure}

Moreover, in Fig.~\ref{fig:denied_stolen}, we consider the case of the unauthorized access with a leaked/stolen token,
where a student obtains the staff token illegally and attempts to use it for resource access.
The student first selects the staff token as the token to present, ``camera1/snapshot'' as the resource (which is exclusive to the staff) and ``GET'' as the action (Image 1 in Fig.~\ref{fig:denied_stolen}).
In the authentication phase, an authentication request for policy ``Division:IS AND {\bf Role:Staff}'' is sent to the owner
(Image 2 in Fig.~\ref{fig:denied_stolen}),
since the selected token is corresponding to the policy ``Division:IS AND {\bf Role:Staff}.''
This makes the owner send back an OTP encrypted using CP-ABE under the policy ``Division:IS AND {\bf Role:Staff}.''
However, the student's private key is associated with the set of attributes \{Division: IS, {\bf Role: Student}\}
and is thus unable to decrypt the encrypted OTP.
Therefore, the student cannot proceed to the access request phase and the attempt to use the token is prevented
(Image 3 in Fig.~\ref{fig:denied_stolen}).
As can be seen here, both the right token and an appropriate private key satisfying the corresponding policy are needed to access resources.

\section{Performance Evaluation}\label{sec:performance_evaluation}
In this section we first evaluate the scalability/throughput performance of our scheme in terms of execution time
and then compare our scheme with DCACI in terms of both single-operation execution time in the case with one subject
and total execution time in the case with multiple subjects.

\subsection{Scalability/Throughput}
We first discuss the scalability/throughput (i.e., the ability of processing access requests per unit time) of our scheme from the point of an ABAC scheme
by measuring the execution time of each operation.

Intuitively, the smaller the execution time is, the higher the scalability will be.
As mentioned in Section \ref{sec:introduction}, our scheme enables flexible and fine-grained access control by properly fining the policies, i.e., setting complicated logic formulas of attributes, which leads to an increased number of attributes contained in the policies.
In addition, in large-scale environments, there can be a large number of policies to describe the complex local authorization policies.
Given these points, we investigate how the execution time changes
1) when the number of attributes in the policies increases
and 2) when the number of policies increases.

\subsubsection{Execution Time vs. Number of Attributes}\label{sec:time_vs_attributes}
As shown in Table~\ref{tab:evaluation_scenario_attributes}, we consider four cases in which the policy consists of 3, 6, 9 and 12 attributes, respectively, and the number of the corresponding access rights was fixed to three.
The average execution time was measured over 100 executions for each operation, using the prototype shown in Fig.~\ref{fig:prototype_configuration}.

\begin{table}[!t]
\caption{Policies used in section \ref{sec:time_vs_attributes}.}
\centering
\includegraphics[clip, width=9cm]{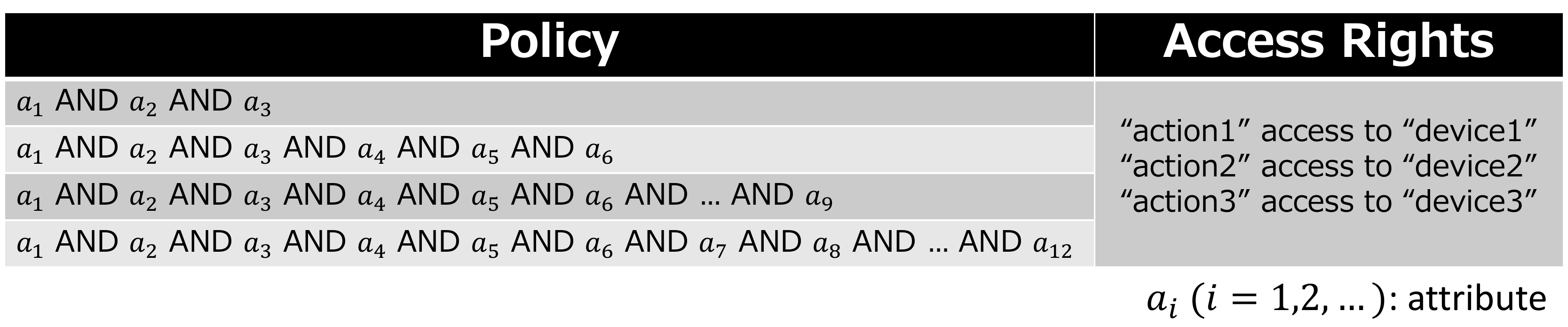}
\label{tab:evaluation_scenario_attributes}
\end{table}

\begin{figure}[!t]
\centering
\includegraphics[clip, width=8cm]{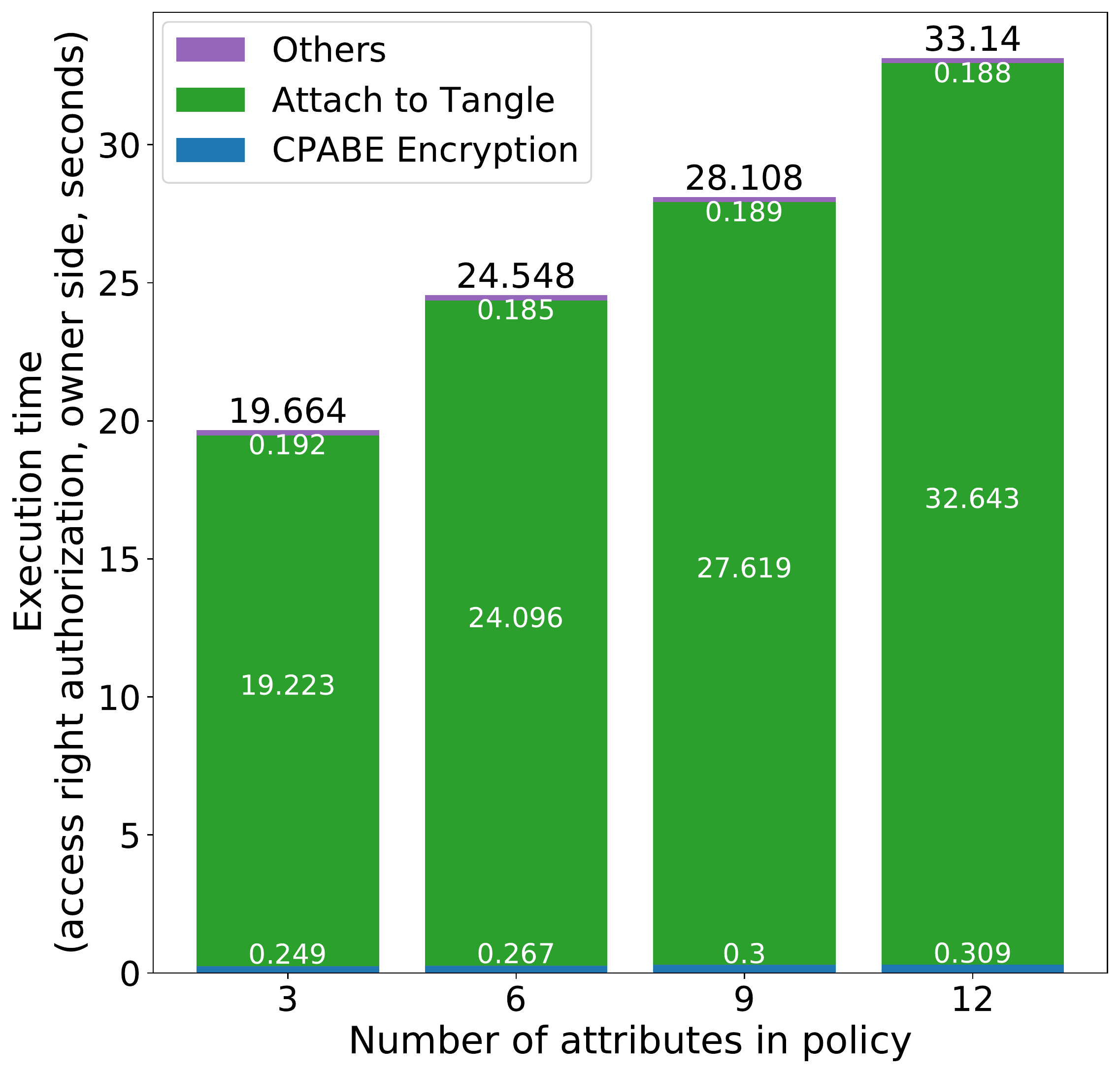}
\caption{Access right authorization time (owner) vs. number of attributes.}
\label{fig:result_attributes_authorization_owner}
\end{figure}

Fig.~\ref{fig:result_attributes_authorization_owner} shows the results for access right authorization at the owner side,
i.e., issuing a token, encrypting it using CP-ABE and attaching it to the Tangle.
We can see that the overall execution time is proportional to the number of attributes in the policy.
A careful observation indicates that attaching the token to the Tangle accounts for the majority of the execution time.
This is mainly because the encrypted token is stored on the Tangle in the form of MAM transactions and we need to perform
time-consuming PoW for each transaction in IOTA.
This part of time is proportional to the number of attributes as well, since more attributes in the policy leads to more transactions being issued.
More specifically, more attributes in the policy leads to a larger encrypted token (this is because the policy is embedded into the ciphertext) and each transaction can contain only a constant amount of data.
Therefore, a large token has to be split into multiple transactions and these transactions are stored together on the Tangle \cite{iotatx}.
It should be noted that the node we used in the experiment ({\tt https://nodes.thetangle.org}) supports remote PoW
(i.e., the node performs the PoW on behalf of the client so that client devices with limited computing power can issue transactions) \cite{iotapow},
and the execution time of attaching the token to the Tangle is thus highly dependent on the condition of the node,
such as computation power and the amount of load at the time of request.
This also means that the execution time is highly dependent on the performance of the client device
when connecting to a node that does not support remote PoW.
As for CP-ABE encryption, it has been shown by the authors in \cite{cpabe} that
the execution time is proportional to the number of attributes in the policy,
and we can see the same results from Fig.~\ref{fig:result_attributes_authorization_owner}.
The remaining part of the time  (i.e., ``Others'' in Fig.~\ref{fig:result_attributes_authorization_owner})
apart from those for attaching the token to the Tangle and CP-ABE encryption
includes time for string operations pertaining to token issuance and conversion between {\tt JSON} objects and Java objects,
which are almost independent of the number of attributes.

\begin{figure}[!t]
\centering
\includegraphics[clip, width=8cm]{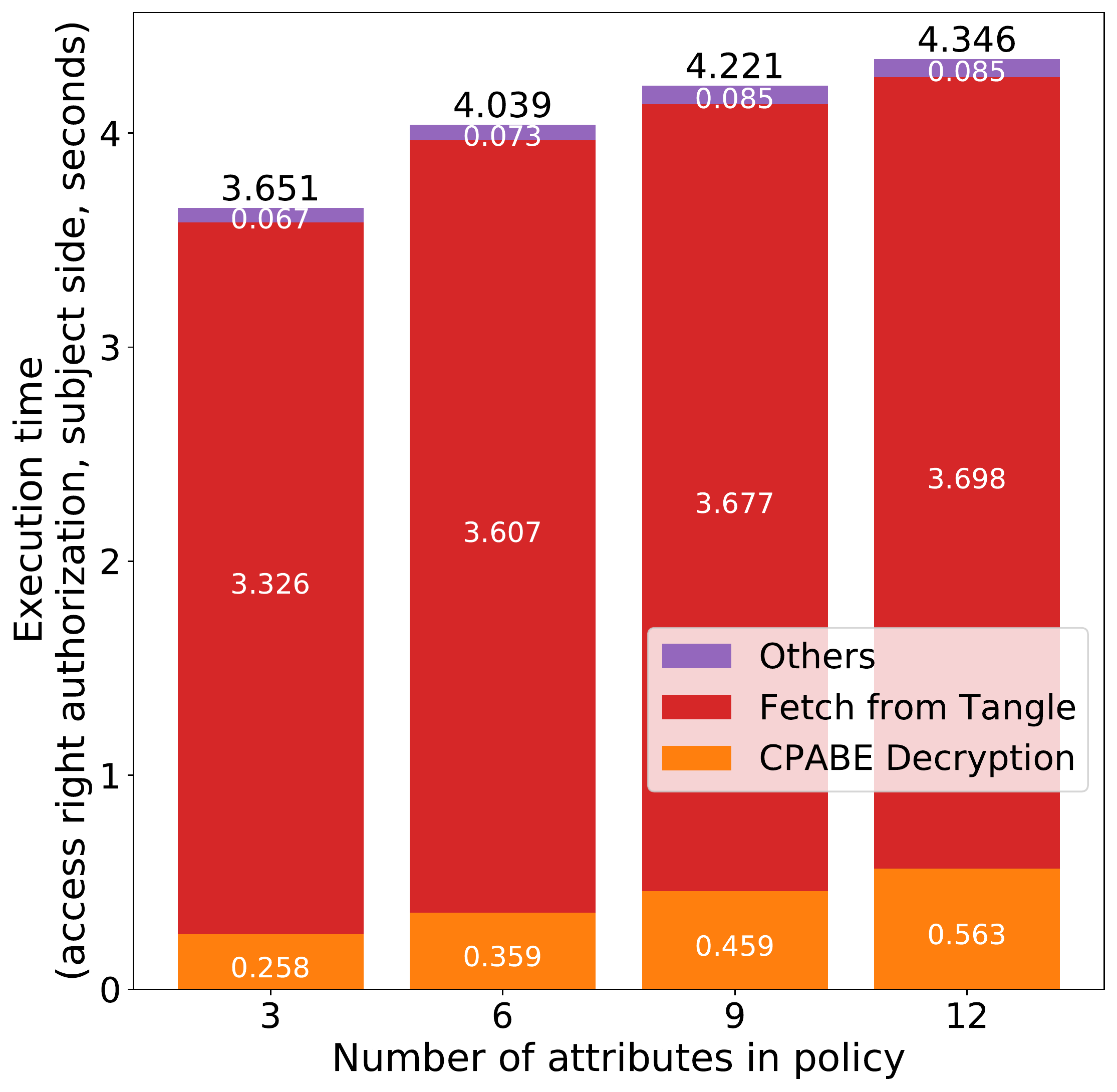}
\caption{Access right authorization time (subject) vs. number of attributes.}
\label{fig:result_attributes_authorization_subject}
\end{figure}

Fig.~\ref{fig:result_attributes_authorization_subject} shows the results for access right authorization at the subject side,
i.e., fetching an encrypted token from the Tangle and decrypting it using CP-ABE.
We can see that the execution time is proportional to the number of attributes in the policy.
We can see from the breakdown of the execution time that fetching the token from the Tangle accounts for the majority of the execution time.
This is because the subject has to fetch all the transactions over which the encrypted token is fragmented,
and then concatenates the transactions to reconstruct the encrypted token (i.e., the inverse operation of attaching the token to the Tangle).
As more attributes lead to more transactions (as mentioned above in the authorization at the owner side),
the execution time is proportional to the number of attributes.
As for CP-ABE decryption, it has been shown by the authors in \cite{cpabe} that the execution time is proportional
to the number of attributes in the policy and the same conclusion can be drawn from Fig.~\ref{fig:result_attributes_authorization_subject}.

\begin{figure}[!t]
\centering
\includegraphics[clip, width=8cm]{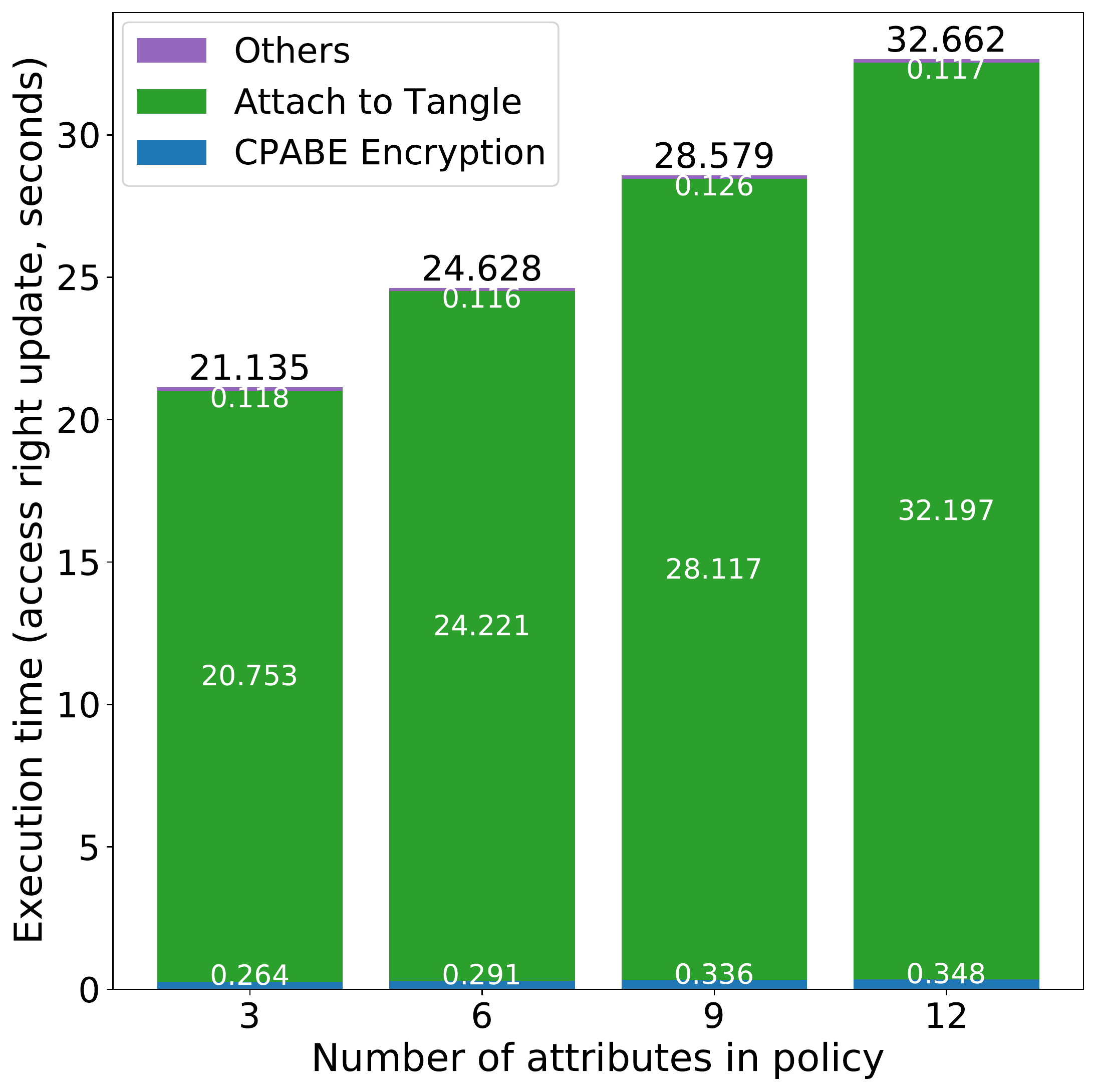}
\caption{Access right update time vs. number of attributes.}
\label{fig:result_attributes_update}
\end{figure}

Fig.~\ref{fig:result_attributes_update} shows the results for access right update,
i.e., issuing a new token, encrypting it using CP-ABE and attaching it to the Tangle by the owner.
The updated access rights were selected randomly from three cases, i.e., reduced rights (one right), increased rights (five rights)
and token inactivation (i.e., setting the status field to ``INACTIVE'').
Since it consists of the same operations as the authorization at the owner side,
we can see similar results from both figures (i.e., Figures  \ref{fig:result_attributes_authorization_owner} and \ref{fig:result_attributes_update}).

\begin{figure}[!t]
\centering
\includegraphics[clip, width=8.8cm]{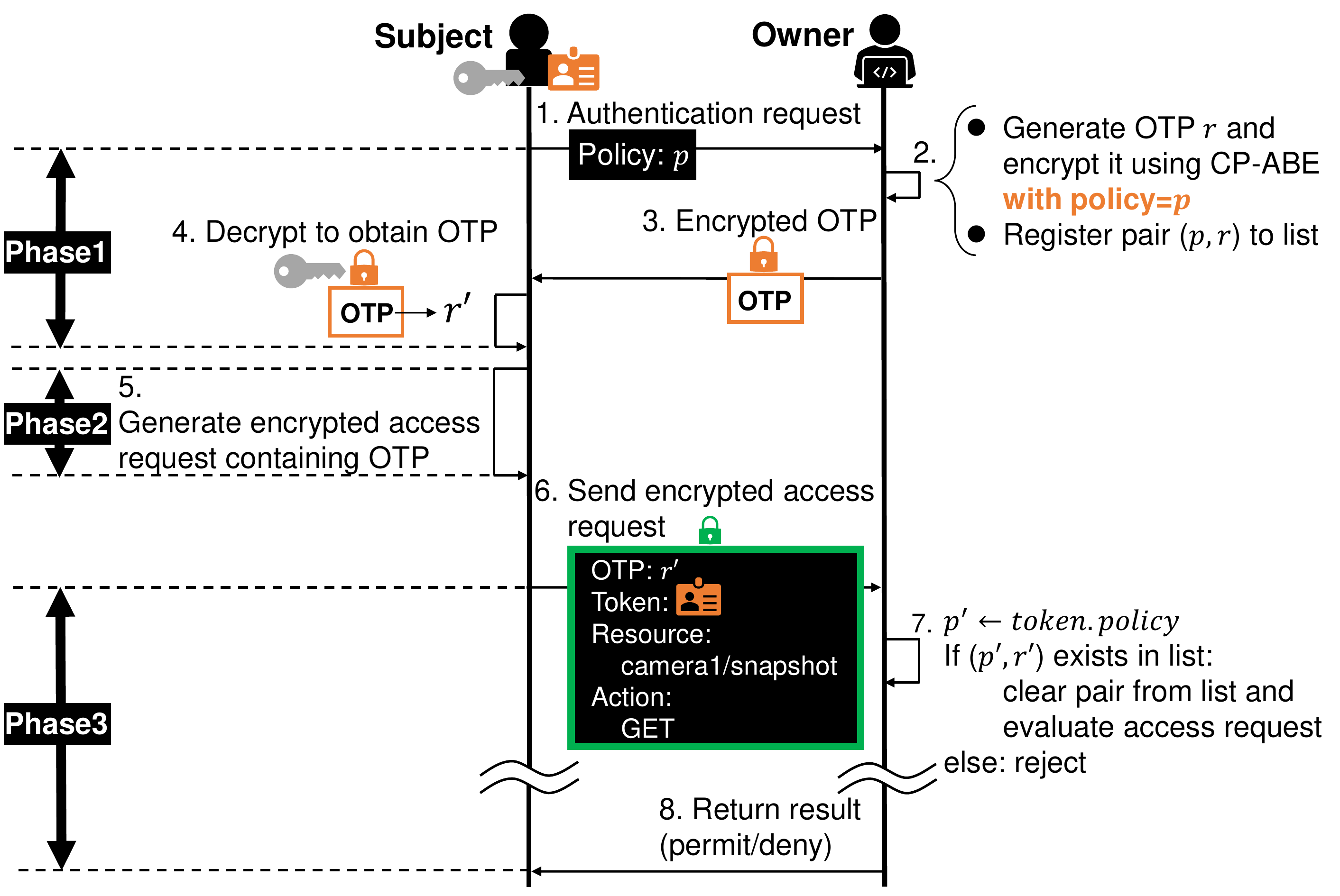}
\vspace{-7mm}
\caption{Overall access right verification process.}
\label{fig:verification_breakdown}
\end{figure}

Next, we evaluate how the number of attributes affects the execution time of access right verification,
which is divided into three phases as shown in Fig.~\ref{fig:verification_breakdown}.

\begin{figure}[!t]
\centering
\includegraphics[clip, width=8cm]{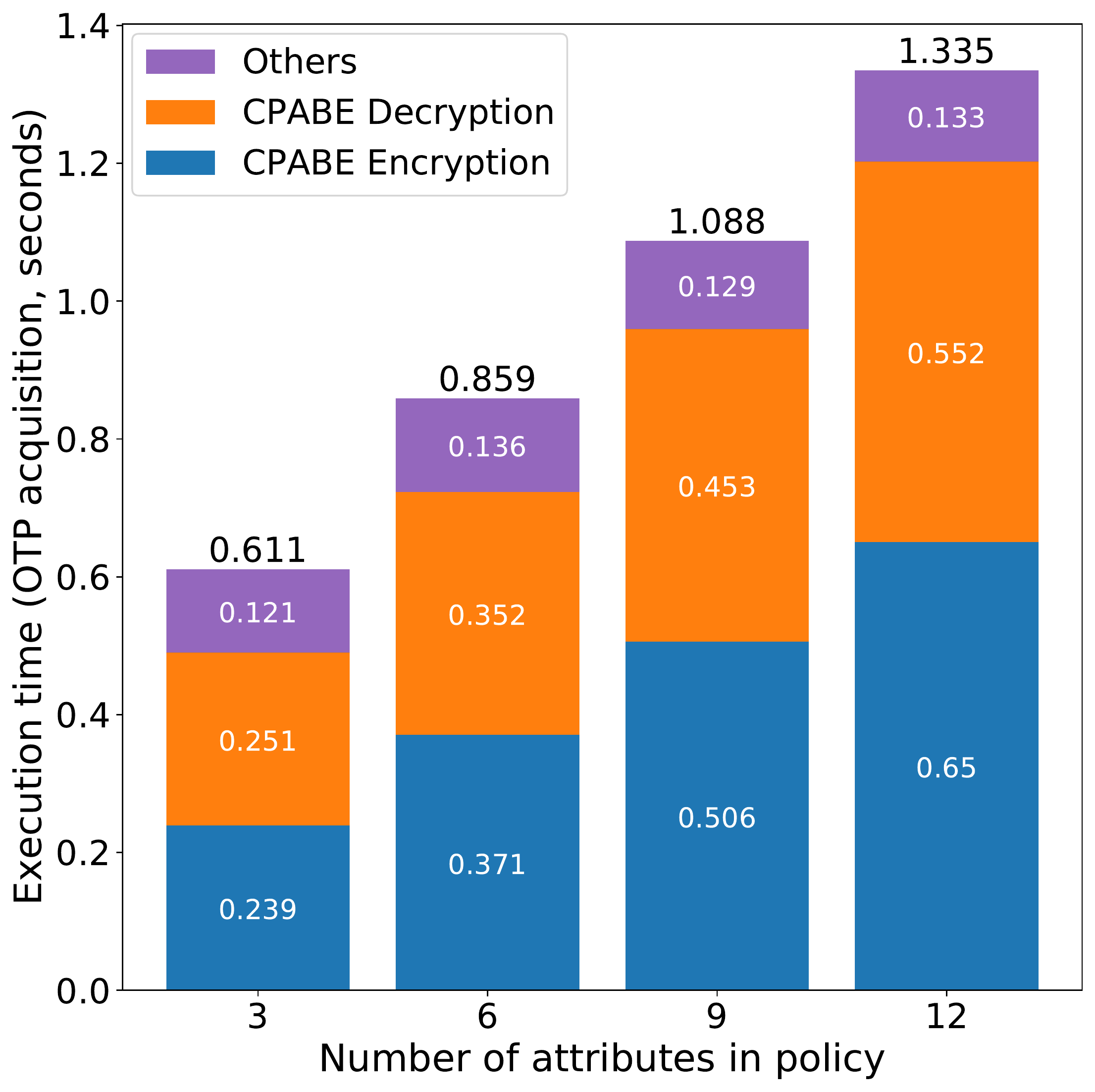}
\caption{Access right verification time (Phase 1) vs. number of attributes.}
\label{fig:result_attributes_otp}
\end{figure}

Fig.~\ref{fig:result_attributes_otp} shows the execution time results for obtaining and decrypting an encrypted OTP from the owner in the first phase.
We can see that the execution time in this phase is proportional to the number of attributes in the policy.
The two main processes, CP-ABE encryption at the owner side (OTP encryption) and
CP-ABE decryption at the subject side (OTP decryption) are both proportional to the number of attributes in the policy, as mentioned earlier.
The remaining time is consumed by other processes including the generation of the authentication request,
simple string operations pertaining to CP-ABE and communication between the subject and the owner.

\begin{figure}[!t]
\centering
\includegraphics[clip, width=8cm]{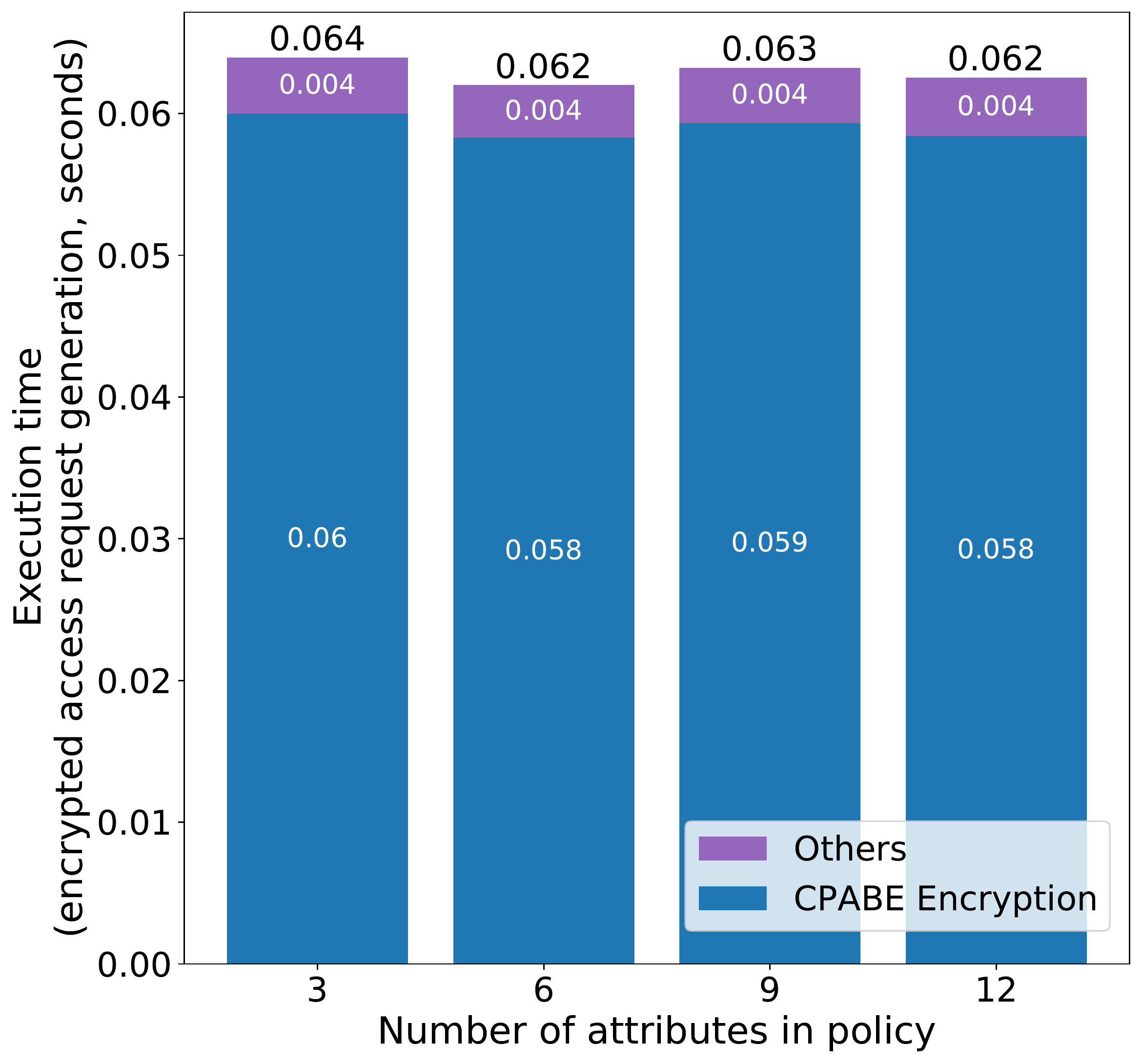}
\caption{Access right verification time (Phase 2) vs. number of attributes.}
\label{fig:result_attributes_request_generation}
\end{figure}

Fig.~\ref{fig:result_attributes_request_generation} shows the average execution time (measured at the subject side)
of the second phase, i.e., generating an encrypted access request using the OTP obtained in the first phase.
We can see that the execution time is almost constant regardless of the number of attributes.
This is because the policy used to encrypt the access request is ``Role: Owner'' (so that only the owner can decrypt the request, as mentioned in Section \ref{sec:proposed_scheme}), which is irrelevant to the ``policy'' (i.e., the policy for encrypting the \emph{token} during authorization) in the x axis.
The ``Others'' corresponds to the time used for simple string operations in CP-ABE.

\begin{figure}[!t]
\centering
\includegraphics[clip, width=8cm]{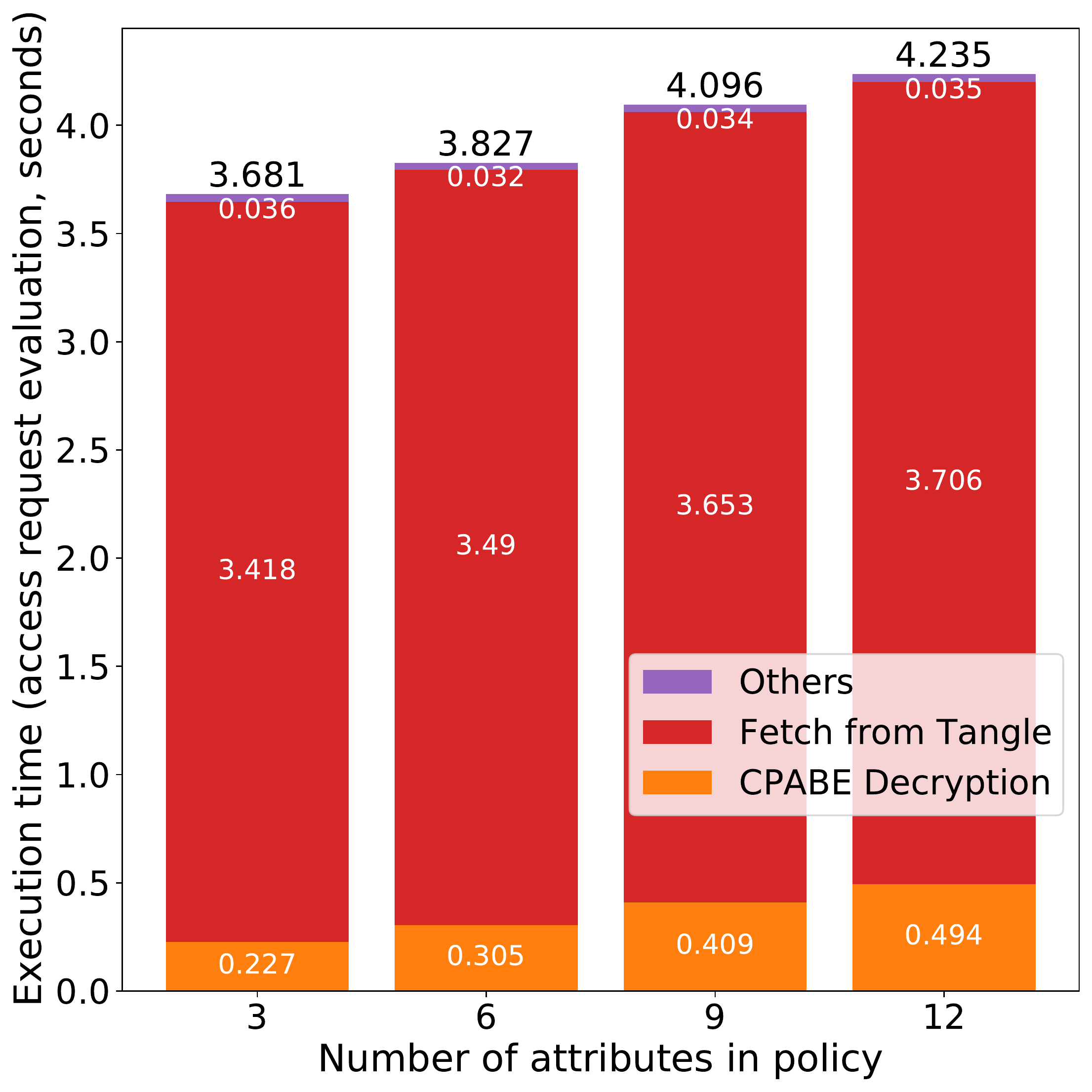}
\caption{Access right verification time (Phase 3) vs. number of attributes.}
\label{fig:result_attributes_request_evaluation}
\end{figure}

Fig.~\ref{fig:result_attributes_request_evaluation} shows the average execution time (owner side) of evaluating the encrypted access request received from the subject in the third phase.
We can see that the execution time is proportional to the number of attributes in the policy.
Similar to the authorization at the subject side, the owner has to fetch the MAM transactions containing the original copy of the presented token,
which accounts for the majority of the execution time and consumes time in proportion to the number of attributes as argued before.
CP-ABE decryption includes decrypting both the access request and the original copy of the token
and consumes time in proportion to the number of attributes, which has also been argued above.

\subsubsection{Execution Time vs. Number of Policies}
In this section we qualitatively analyze how the number of policies affect the execution time of each operation.

\begin{itemize}
\item Access Right Authorization
	\begin{itemize}
		\item Owner Side \\
		Since tokens are issued and recorded to the Tangle for each policy,
		the execution time of access right authorization is proportional to the number of policies and is equal to the sum of the execution time
		of issuing and recording the tokens for all policies,
		each of which depends on the number of attributes included in the policy as shown in Section \ref{sec:time_vs_attributes}.
		Although this is expected to take a considerable time due to attaching the tokens to the Tangle,
		it is done only once because all subjects can be authorized once all the tokens are recorded to the Tangle.

		\item Subject Side \\
		As introduced in Section \ref{sec:proposed_scheme}, a MAM channel is generated for each policy
		when the encrypted tokens are recorded to the Tangle.
		Subjects can directly refer to the channels of their interests using the addresses associated with the tokens.
		Therefore, obtaining tokens from the Tangle is independent of other policies and thus the number of policies does not affect the execution time.
	\end{itemize}

\item Access Right Update \\
	As introduced in Section \ref{sec:proposed_scheme}, the tokens of the same policy are linked together in a MAM channel.
	The owner simply attaches the new token to the corresponding channel, without any information about other policies.
	Therefore, token update can be performed individually and is independent of the number of policies.

\item Access Right Verification \\
	We discuss each phase illustrated in Fig.~\ref{fig:verification_breakdown}.
	\begin{itemize}
		\item OTP Acquisition (Phase 1 in Fig.~\ref{fig:verification_breakdown}) \\
			When making an authentication request, the subject can extract the policy from his/her token and information about other policies is unnecessary.
			CP-ABE encryption and decryption, which are the main operations, do not require other policies either.
			Therefore, the increase of the number of policies does not affect the time of OTP acquisition.

		\item Encrypted Access Request Generation (Phase 2 in Fig.~\ref{fig:verification_breakdown})  \\
			The information needed to generate an encrypted access request is the OTP obtained through the authentication request
			(i.e., in the OTP acquisition phase), the token, the resource to access and
			the action to perform, all of which are independent of other policies.
			Therefore, the increase of the number of policies does not affect the time of access request generation.

		\item Access Request Evaluation (Phase 3 in Fig.~\ref{fig:verification_breakdown}) \\
			The main operation here is fetching the original copy of the presented token.
			Similar to the authorization process at the subject side, this can be performed independently thanks to MAM channels and is
			thus irrelevant to the number of policies.
	\end{itemize}

\end{itemize}

As argued above, only the initial token recording involves iterating over all policies
and operations that are invoked frequently (e.g., fetching encrypted tokens from the Tangle) are independent of the number of policies.
Therefore, we can conclude that the increase of the number of policies has limited impact on the overall execution time.

\subsection{Comparison with DCACI}
In this subsection we compare our scheme with DCACI in terms of both single-operation execution time and total execution time.
For comparison we have implemented the DCACI scheme based on the paper \cite{dcaci}, using the IOTA Mainnet.
The prototype shown in Fig.~\ref{fig:prototype_configuration} was used for both schemes, except that the subject side was executed
on the Surface Laptop 3 for DCACI.
As for the evaluation scenario we suppose an educational institution consisting of students and staff members,
and the access rights will be authorized to them according to Table~\ref{tab:local_policy}.

We consider two scenarios:
1) one object owner and one subject (one-to-one scenario) for comparing single-operation execution time
and
2) one object owner and multiple subjects (one-to-many scenario) for comparing the total execution time.
We show that our scheme can greatly reduce the execution time of authorizing access rights to a large number of subjects by one-to-many access control.

\subsubsection{One Owner and One Subject (One-to-One Scenario)}
We first consider one owner and one subject who is a member of the staff.
The average execution time for each operation has been measured for both schemes and compared.

\begin{figure}[!t]
\centering
\includegraphics[clip, width=8cm]{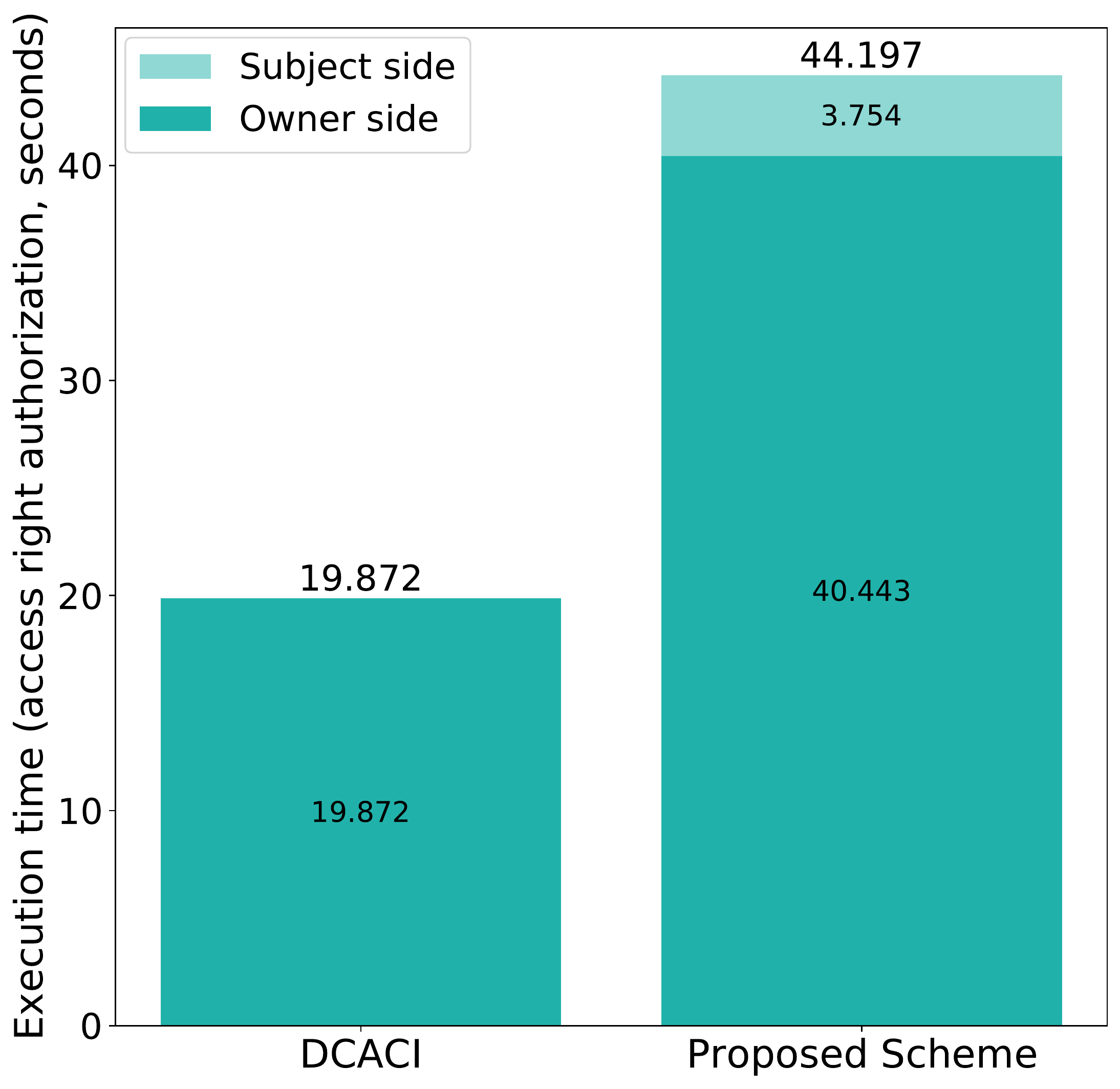}
\caption{Comparison with DCACI: access right authorization.}
\label{fig:result_comparison_auth}
\end{figure}

Fig.~\ref{fig:result_comparison_auth} shows the results for access right authorization.
As can be seen in the figure, our scheme requires extra time, especially at the subject side.
This is because the subject has to fetch and decrypt the encrypted token from the Tangle in our scheme,
while in DCACI the owner directly issues the token to the subject in response to the token issuance request of the subject.

\begin{figure}[!t]
\centering
\includegraphics[clip, width=8cm]{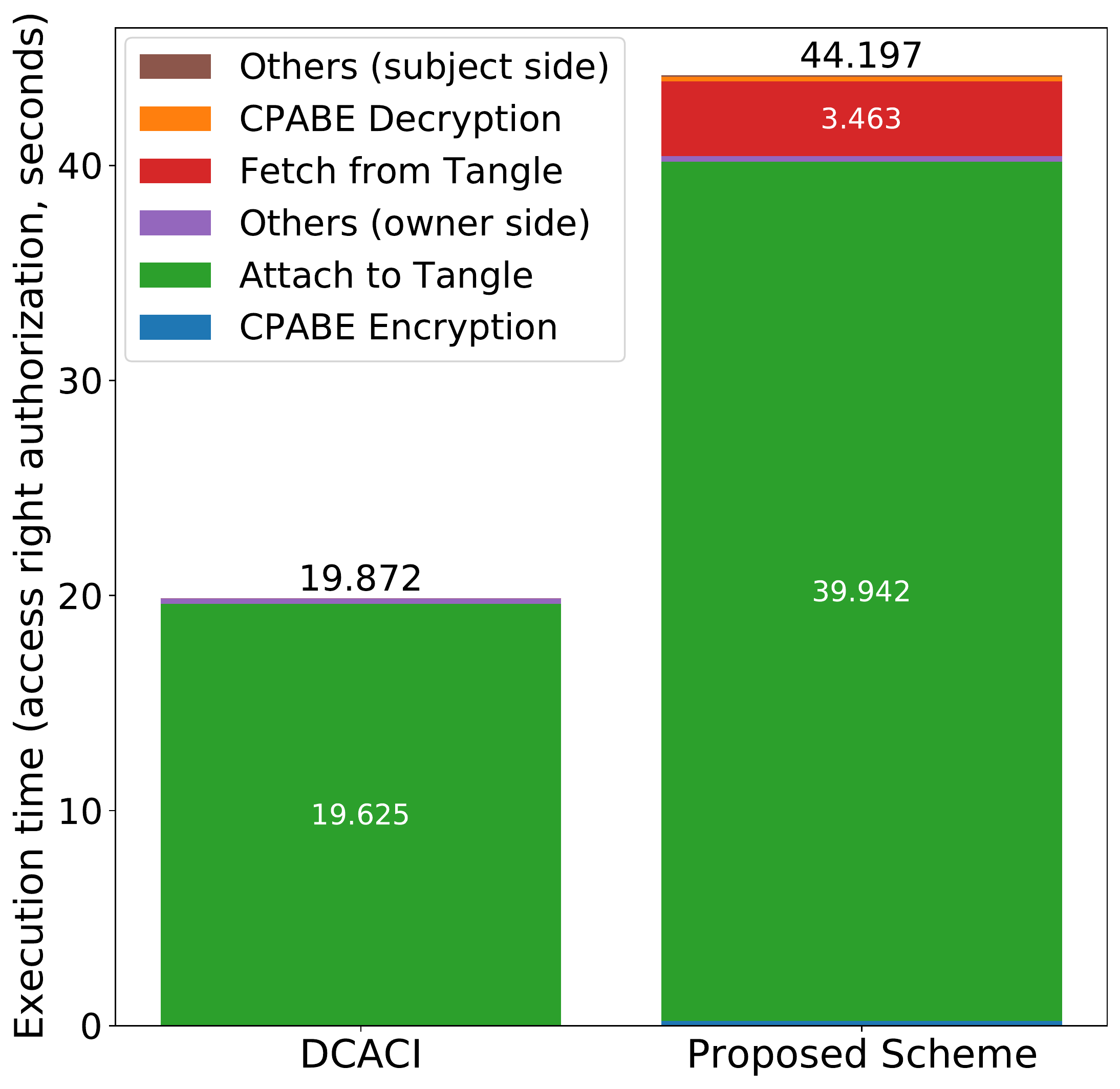}
\caption{Comparison with DCACI: access right authorization (with breakdown).}
\label{fig:result_comparison_auth_breakdown}
\end{figure}

Moreover, the breakdown of the execution time is shown in Fig.~\ref{fig:result_comparison_auth_breakdown}.
We can see that attaching the token to the Tangle accounts for the majority of the execution time, as also discussed previously.
Attaching the token to the Tangle takes longer in our scheme, because the token to attach in our scheme is larger than that in DCACI due to CP-ABE encryption.
Although the increased token size also leads to an increased CP-ABE encryption time in our scheme compared with DCACI,
this increase is much smaller than that caused by attaching the token to the Tangle.
Therefore, we can conclude that CP-ABE encryption has limited impact on the overall execution time.
Same phenomenons can be observed at the subject side, where fetching the token from the Tangle consumes most of the time.

\begin{figure}[!t]
\centering
\includegraphics[clip, width=8cm]{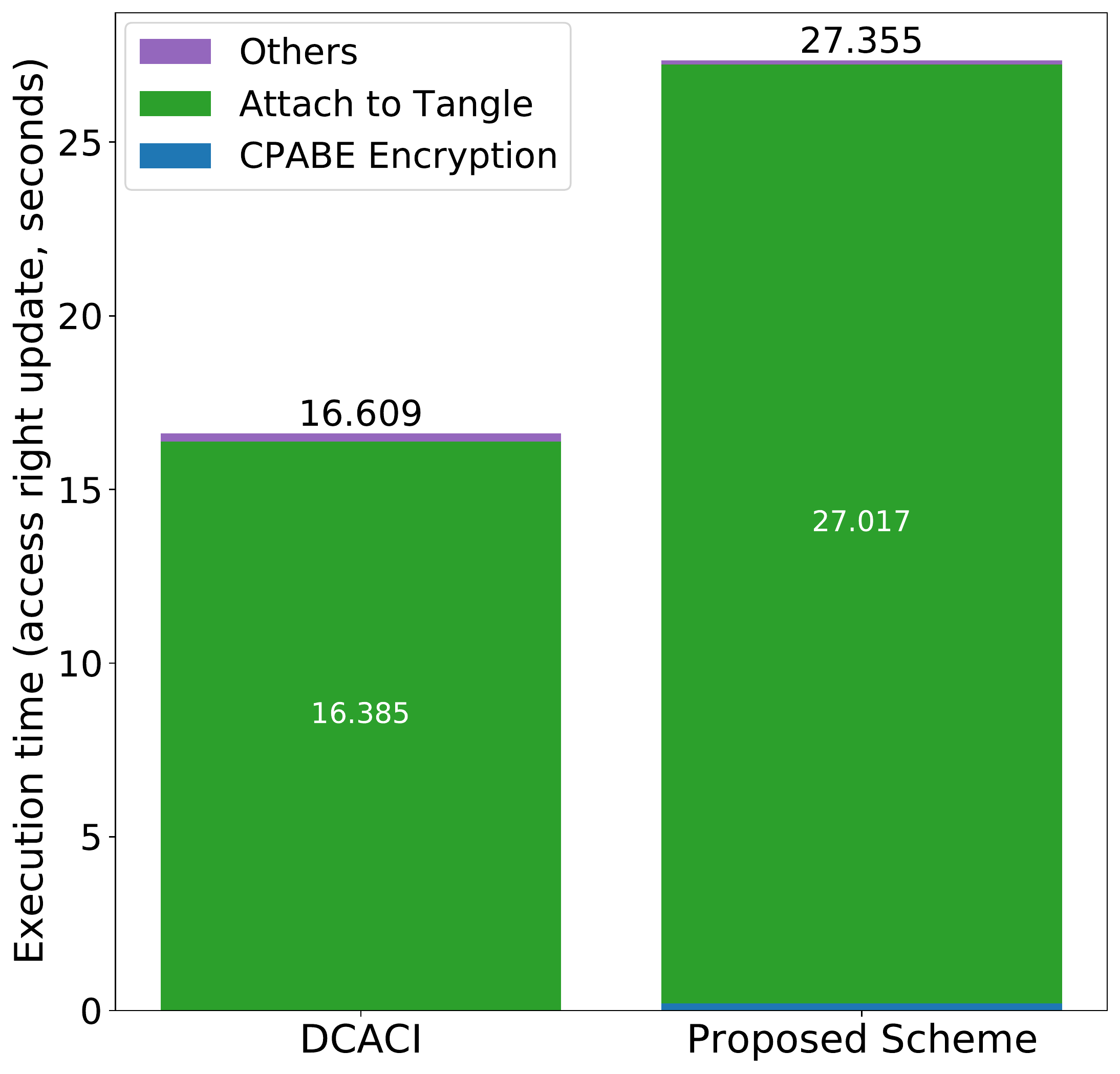}
\caption{Comparison with DCACI: access right update.}
\label{fig:result_comparison_update}
\end{figure}

Fig.~\ref{fig:result_comparison_update} shows the comparison results for access right update.
The updated access rights were selected randomly from three cases, i.e., adding rights, reducing rights and setting status to ``INACTIVE.''
Similar to the results in the authorization, attaching the token to the Tangle is dominant and CP-ABE encryption has little impact.

\begin{figure}[!t]
\centering
\includegraphics[clip, width=8cm]{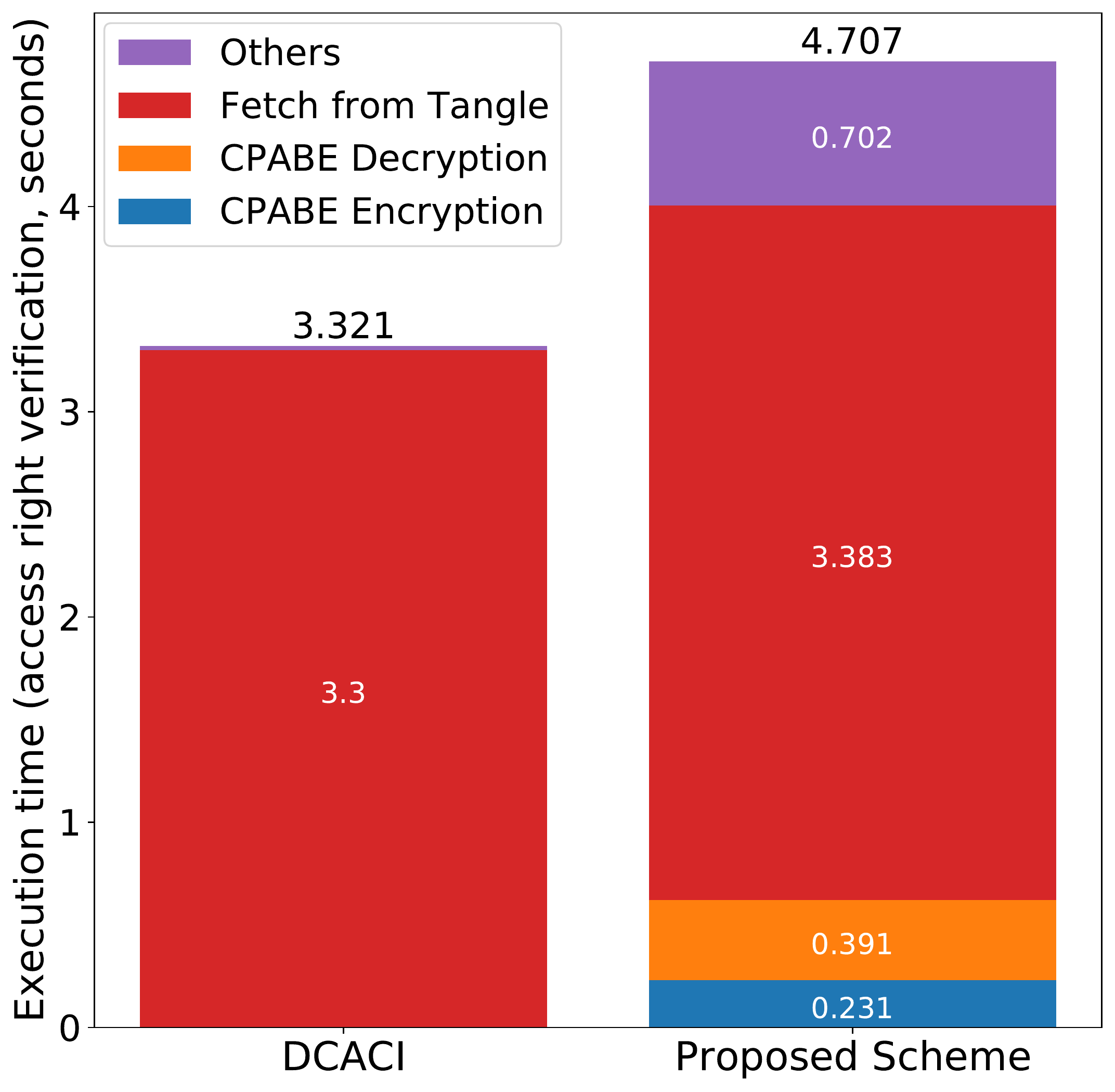}
\caption{Comparison with DCACI: access right verification.}
\label{fig:result_comparison_ver}
\end{figure}

Finally, Fig.~\ref{fig:result_comparison_ver} shows the comparison results for access right verification.
We can see that fetching the token from the Tangle accounts for the majority of the execution time.
In the proposed scheme, the ``Others'' takes a longer time than that of DCACI,
because the subject and owner have to communicate two times,
i.e., one in the authentication phase and the other in the request phase.

As can be seen above, experimental results show that our scheme is slower than DCACI for every operation.
However, it should be noted that the results for DCACI do not include the security measures,
while the results for our scheme do (e.g., establishing secure channels and authentication of subjects).

\subsubsection{One Owner and Multiple Subjects (One-to-Many Scenario)}
Taking into account that our scheme supports one-to-many access control, i.e., one token can be used to authorize multiple subjects,
we can expect that our scheme outperforms DCACI when authorizing a large number of subjects.
To demonstrate this, we consider the case in which the owner authorizes multiple subjects
(more specifically, 1200 subjects consisting of 1000 students and 200 staff members),
and compare the \emph{total} execution time of authorizing \emph{all} subjects.

\begin{figure}[!t]
\centering
\includegraphics[clip, width=8cm]{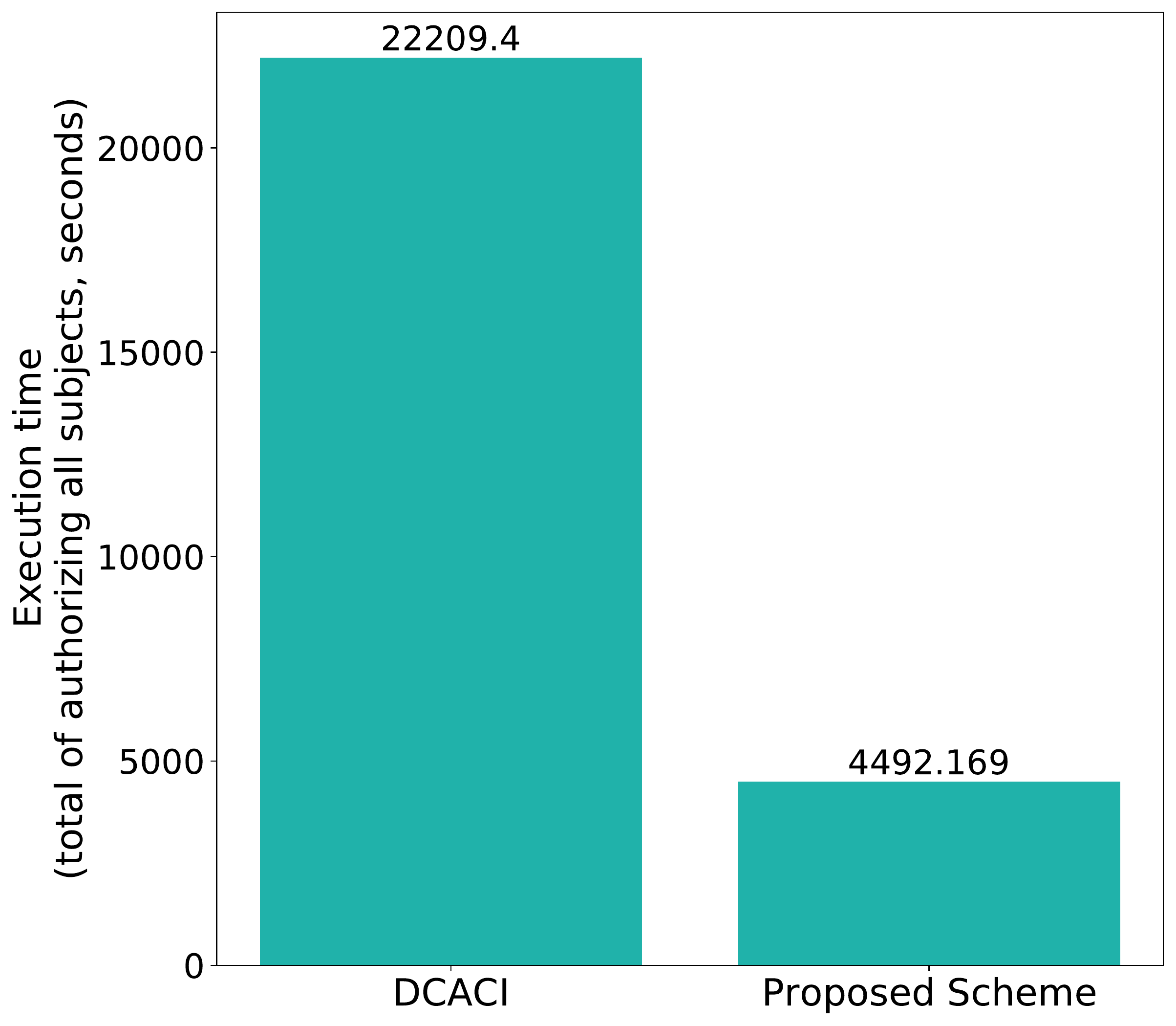}
\caption{Comparison with DCACI: total execution time of authorizing $1000$ students and $200$ staff members.}
\label{fig:result_comparison_onetomany}
\end{figure}

Table \ref{tab:enum_operations} shows the comparison of the operations needed to authorize all the subjects,
and Table \ref{tab:enum_avg} shows the average execution time of each operation over 100 executions.
Based on this, the total execution time is calculated and compared in Fig.~\ref{fig:result_comparison_onetomany}.
We can see that our scheme can authorize the subjects in a significantly shorter time, as expected.
This is because the number of times the owner has to attach data to the Tangle is much less in our scheme,
which is equal to the number of policies.
Since there are two policies in this scenario, the owner attaches encrypted tokens to the Tangle for a mere two times,
while in DCACI original copies of tokens are attached to the Tangle for every subject, i.e., 1200 times.
Although fetching tokens from the Tangle has to be performed by every subject in our scheme,
it takes much less time than attaching tokens to the Tangle, leading to the overall reduction in execution time.

\begin{table}[t]
  \begin{center}
    \caption{Operations to authorize $1000$ students and $200$ staff members.}
    \label{tab:enum_operations}
    \vspace{5mm}
    \begin{tabular}{|l|r|} \hline
       Scheme & Operations \\ \hline \hline
      DCACI &
      \begin{tabular}{c}
      	$1000 \times \mathit{GrantAccess} $ to student
      	\\ $+200 \times \mathit{GrantAccess} $ to staff member
      \end{tabular}\\ \hline
      Proposed scheme &
      \begin{tabular}{c}
      	$1 \times $ publish student token
      	\\ $+1 \times $ publish staff token
      	\\ $+1000 \times $ obtain token by student
      	\\ $+200 \times $ obtain token by staff member
      \end{tabular}\\ \hline
    \end{tabular}
  \end{center}
\end{table}

\begin{table}[t]
  \begin{center}
    \caption{Comparison with DCACI: average execution time of each operation.}
    \label{tab:enum_avg}
    \vspace{5mm}
    \begin{tabular}{|l|r|} \hline
      Operation &
      \begin{tabular}{c}
      	Average execution time
      	\\ (seconds)
      \end{tabular} \\ \hline \hline
      \emph{GrantAccess} to student (DCACI) & $18.235$ \\ \hline
      \emph{GrantAccess} to staff member (DCACI) & $19.872$ \\ \hline
      Publish student token (proposed scheme) & $32.926$ \\ \hline
      Publish staff token (proposed scheme) & $40.443$ \\ \hline
      Obtain student token (proposed scheme) & $3.668$ \\ \hline
      Obtain staff token (proposed scheme) & $3.754$ \\ \hline
    \end{tabular}
  \end{center}
\end{table}

As can be seen here, our scheme provides faster authorization for a large number of subjects than DCACI.
In addition, the number of operations object owners have to perform is greatly reduced, alleviating the burden of the owners.

\section{Conclusions}\label{sec:conclusions}
In this paper we have proposed an IOTA-based access control framework in which Attribute-Based Access Control (ABAC) and
Capability-Based Access Control (CapBAC) are combined
by leveraging the Ciphertext-Policy Attribute-Based Encryption (CP-ABE) technology.
Thanks to CP-ABE, our scheme overcomes the drawbacks that exist in the previous framework named
Decentralized Capability-Based Access Control framework using IOTA (DCACI)
and provides more secure, fine-grained and scalable access control.
We have shown the feasibility of our scheme by implementing a practical proof-of-concept prototype using the IOTA Mainnet and IoT devices.
We have also evaluated the performance of our scheme in terms of execution time and compared it with DCACI.
Experimental results show that our scheme enables object owners to authorize access rights to a large number of subjects
in a much shorter time than the DCACI scheme.
Furthermore, we have discussed the scalability of our scheme by considering an increased number of attributes and policies.
The results show that the execution time for each operation is proportional to the number of attributes involved,
and that the number of policies has little effect on the overall execution time for most operations.

\bibliographystyle{IEEEtran}
\bibliography{IEEEabrv,references}

\begin{thebibliography}{10}
\providecommand{\url}[1]{#1}
\csname url@samestyle\endcsname
\providecommand{\newblock}{\relax}
\providecommand{\bibinfo}[2]{#2}
\providecommand{\BIBentrySTDinterwordspacing}{\spaceskip=0pt\relax}
\providecommand{\BIBentryALTinterwordstretchfactor}{4}
\providecommand{\BIBentryALTinterwordspacing}{\spaceskip=\fontdimen2\font plus
\BIBentryALTinterwordstretchfactor\fontdimen3\font minus
  \fontdimen4\font\relax}
\providecommand{\BIBforeignlanguage}[2]{{%
\expandafter\ifx\csname l@#1\endcsname\relax
\typeout{** WARNING: IEEEtran.bst: No hyphenation pattern has been}%
\typeout{** loaded for the language `#1'. Using the pattern for}%
\typeout{** the default language instead.}%
\else
\language=\csname l@#1\endcsname
\fi
#2}}
\providecommand{\BIBdecl}{\relax}
\BIBdecl

\bibitem{nakanishi2020iota}
R.~Nakanishi, Y.~Zhang, M.~Sasabe, and S.~Kasahara, ``{IOTA}-based access
  control framework for the {Internet} of things,'' in \emph{Proc. 2nd
  Conference on Blockchain Research and Applications for Innovative Networks
  and Services (BRAINS)}, 2020, pp. 87--95.

\bibitem{gartner}
\BIBentryALTinterwordspacing
Gartner identifies top 10 strategic {IoT} technologies and trends. [Online].
  Available:
  \url{https://www.gartner.com/en/newsroom/press-releases/2018-11-07-gartner-identifies-top-10-strategic-iot-technologies-and-trends/.}
\BIBentrySTDinterwordspacing

\bibitem{ben2019internet}
M.~Ben-Daya, E.~Hassini, and Z.~Bahroun, ``Internet of things and supply chain
  management: A literature review,'' \emph{International Journal of Production
  Research}, vol.~57, pp. 4719--4742, 2019.

\bibitem{qadri2020future}
Y.~A. Qadri, A.~Nauman, Y.~B. Zikria, A.~V. Vasilakos, and S.~W. Kim, ``The
  future of healthcare {Internet} of things: A survey of emerging
  technologies,'' \emph{{IEEE} Commun. Surveys Tuts.}, vol.~22, pp. 1121--1167,
  2020.

\bibitem{yang2019internet}
H.~Yang, S.~Kumara, S.~T. Bukkapatnam, and F.~Tsung, ``The {Internet} of things
  for smart manufacturing: A review,'' \emph{IISE Transactions}, vol.~51, pp.
  1190--1216, 2019.

\bibitem{haddadpajouh2019survey}
H.~HaddadPajouh, A.~Dehghantanha, R.~M. Parizi, M.~Aledhari, and H.~Karimipour,
  ``A survey on {Internet} of things security: Requirements, challenges, and
  solutions,'' \emph{Internet of Things}, 2019.

\bibitem{ande2020internet}
R.~Ande, B.~Adebisi, M.~Hammoudeh, and J.~Saleem, ``Internet of things:
  Evolution and technologies from a security perspective,'' \emph{Sustainable
  Cities and Society}, vol.~54, 2020.

\bibitem{butun2019security}
I.~Butun, P.~{\"O}sterberg, and H.~Song, ``Security of the {Internet} of
  things: Vulnerabilities, attacks, and countermeasures,'' \emph{{IEEE} Commun.
  Surveys Tuts.}, vol.~22, pp. 616--644, 2019.

\bibitem{attack}
N.~Neshenko, E.~Bou-Harb, J.~Crichigno, G.~Kaddoum, and N.~Ghani,
  ``Demystifying {IoT} security: An exhaustive survey on {IoT} vulnerabilities
  and a first empirical look on {Internet-scale} {IoT} exploitations,''
  \emph{{IEEE} Commun. Surveys Tuts.}, vol.~21, pp. 2702--2733, 2019.

\bibitem{xu2018blendcac}
R.~Xu, Y.~Chen, E.~Blasch, and G.~Chen, ``{BlendCAC}: A blockchain-enabled
  decentralized capability-based access control for {IoTs},'' in \emph{Proc.
  2018 IEEE International Conference on Internet of Things (iThings) and IEEE
  Green Computing and Communications (GreenCom) and IEEE Cyber, Physical and
  Social Computing (CPSCom) and IEEE Smart Data (SmartData)}, 2018, pp.
  1027--1034.

\bibitem{xu2019exploration}
R.~Xu, Y.~Chen, E.~Blasch, and G.~Chen, ``Exploration of blockchain-enabled
  decentralized capability-based access control strategy for space situation
  awareness,'' \emph{Optical Engineering}, vol.~58, 2019.

\bibitem{nakamura2019capbac}
Y.~Nakamura, Y.~Zhang, M.~Sasabe, and S.~Kasahara, ``Capability-based access
  control for the {Internet} of things: An {Ethereum} blockchain-based
  scheme,'' in \emph{Proc. IEEE GLOBECOM 2019}, 2019.

\bibitem{nakamura2020exploiting}
Y.~Nakamura, Y.~Zhang, M.~Sasabe, and S.~Kasahara, ``Exploiting smart contracts
  for capability-based access control in the {Internet} of things,''
  \emph{Sensors}, vol.~20, 2020.

\bibitem{dukkipati2018decentralized}
C.~Dukkipati, Y.~Zhang, and L.~C. Cheng, ``Decentralized, blockchain based
  access control framework for the heterogeneous {Internet} of things,'' in
  \emph{Proc. 3rd ACM Workshop on Attribute-Based Access Control}, 2018, pp.
  61--69.

\bibitem{maesa2019blockchain}
D.~D.~F. Maesa, P.~Mori, and L.~Ricci, ``A blockchain based approach for the
  definition of auditable access control systems,'' \emph{Computers \&
  Security}, vol.~84, pp. 93--119, 2019.

\bibitem{yutaka2019abac}
M.~Yutaka, Y.~Zhang, M.~Sasabe, and S.~Kasahara, ``Using {Ethereum} blockchain
  for distributed attribute-based access control in the {Internet} of things,''
  in \emph{Proc. IEEE GLOBECOM 2019}, 2019.

\bibitem{zhang2020attribute}
Y.~Zhang, M.~Yutaka, M.~Sasabe, and S.~Kasahara, ``Attribute-based access
  control for smart cities: A smart contract-driven framework,'' \emph{{IEEE}
  Internet Things J.}, 2020.

\bibitem{cruz2018rbac}
J.~P. Cruz, Y.~Kaji, and N.~Yanai, ``{RBAC-SC}: Role-based access control using
  smart contract,'' \emph{{IEEE} Access}, vol.~6, pp. 12\,240--12\,251, 2018.

\bibitem{rahman2020context}
M.~U. Rahman, B.~Guidi, F.~Baiardi, and L.~Ricci, ``Context-aware and dynamic
  role-based access control using blockchain,'' in \emph{Proc. International
  Conference on Advanced Information Networking and Applications}, 2020, pp.
  1449--1460.

\bibitem{zhang2019IoT}
Y.~{Zhang}, S.~{Kasahara}, Y.~{Shen}, X.~{Jiang}, and J.~{Wan}, ``Smart
  contract-based access control for the {Internet} of things,'' \emph{{IEEE}
  Internet Things J.}, vol.~6, pp. 1594--1605, 2019.

\bibitem{sultana2020data}
T.~Sultana, A.~Almogren, M.~Akbar, M.~Zuair, I.~Ullah, and N.~Javaid, ``Data
  sharing system integrating access control mechanism using blockchain-based
  smart contracts for {IoT} devices,'' \emph{Applied Sciences}, vol.~10, 2020.

\bibitem{novo2018blockchain}
O.~Novo, ``Blockchain meets {IoT}: An architecture for scalable access
  management in {IoT},'' \emph{{IEEE} Internet Things J.}, vol.~5, pp.
  1184--1195, 2018.

\bibitem{ouaddah2016fairaccess}
A.~Ouaddah, A.~Abou~Elkalam, and A.~Ait~Ouahman, ``{FairAccess}: A new
  blockchain-based access control framework for the {Internet} of things,''
  \emph{Security and Communication Networks}, vol.~9, pp. 5943--5964, 2016.

\bibitem{maesa2017blockchain}
D.~D.~F. Maesa, P.~Mori, and L.~Ricci, ``Blockchain based access control,'' in
  \emph{Proc. IFIP international conference on distributed applications and
  interoperable systems}, 2017, pp. 206--220.

\bibitem{pinno2017controlchain}
O.~J.~A. Pinno, A.~R.~A. Gregio, and L.~C. De~Bona, ``{ControlChain}:
  Blockchain as a central enabler for access control authorizations in the
  {IoT},'' in \emph{Proc. IEEE GLOBECOM 2017}, 2017.

\bibitem{ding2019novel}
S.~Ding, J.~Cao, C.~Li, K.~Fan, and H.~Li, ``A novel attribute-based access
  control scheme using blockchain for {IoT},'' \emph{{IEEE} Access}, vol.~7,
  pp. 38\,431--38\,441, 2019.

\bibitem{zhu2018tbac}
Y.~Zhu, Y.~Qin, G.~Gan, Y.~Shuai, and W.~C.-C. Chu, ``{TBAC}:
  {Transaction}-based access control on blockchain for resource sharing with
  cryptographically decentralized authorization,'' in \emph{Proc. 2018 IEEE
  42nd Annual Computer Software and Applications Conference (COMPSAC)}, vol.~1,
  2018, pp. 535--544.

\bibitem{btc}
\BIBentryALTinterwordspacing
Bitcoin -- open source {P2P} money. [Online]. Available:
  \url{https://bitcoin.org/en/.}
\BIBentrySTDinterwordspacing

\bibitem{eth}
\BIBentryALTinterwordspacing
Home | {Ethereum}. [Online]. Available: \url{https://ethereum.org/.}
\BIBentrySTDinterwordspacing

\bibitem{ethsmartcontract}
\BIBentryALTinterwordspacing
Introduction to smart contracts. [Online]. Available:
  \url{https://ethereum.org/en/developers/docs/smart-contracts/.}
\BIBentrySTDinterwordspacing

\bibitem{nist2018blockchain}
\BIBentryALTinterwordspacing
Blockchain technology overview. [Online]. Available:
  \url{https://nvlpubs.nist.gov/nistpubs/ir/2018/NIST.IR.8202.pdf.}
\BIBentrySTDinterwordspacing

\bibitem{conoscenti2016blockchain}
M.~Conoscenti, A.~Vetro, and J.~C. De~Martin, ``Blockchain for the {Internet}
  of things: A systematic literature review,'' in \emph{Proc. 2016 IEEE/ACS
  13th International Conference of Computer Systems and Applications (AICCSA)},
  2016.

\bibitem{iota}
\BIBentryALTinterwordspacing
The next generation of distributed ledger technology | {IOTA}. [Online].
  Available: \url{https://www.iota.org/.}
\BIBentrySTDinterwordspacing

\bibitem{dcaci}
S.~K. {Pinjala} and K.~M. {Sivalingam}, ``{DCACI}: A decentralized lightweight
  capability based access control framework using {IOTA} for {Internet} of
  things,'' in \emph{Proc. 2019 IEEE 5th World Forum on Internet of Things
  (WF-IoT)}, 2019, pp. 13--18.

\bibitem{cpabe}
J.~{Bethencourt}, A.~{Sahai}, and B.~{Waters}, ``Ciphertext-policy
  attribute-based encryption,'' in \emph{Proc. IEEE Symposium on Security and
  Privacy (SP '07)}, 2007, pp. 321--334.

\bibitem{sandhu1994access}
R.~S. Sandhu and P.~Samarati, ``Access control: Principle and practice,''
  \emph{{IEEE} Commun. Mag.}, vol.~32, pp. 40--48, 1994.

\bibitem{sandhu1996role}
R.~S. Sandhu, E.~J. Coyne, H.~L. Feinstein, and C.~E. Youman, ``Role-based
  access control models,'' \emph{Computer}, vol.~29, pp. 38--47, 1996.

\bibitem{hu2015attribute}
V.~C. Hu, D.~R. Kuhn, D.~F. Ferraiolo, and J.~Voas, ``Attribute-based access
  control,'' \emph{Computer}, vol.~48, pp. 85--88, 2015.

\bibitem{gusmeroli2013capability}
S.~Gusmeroli, S.~Piccione, and D.~Rotondi, ``A capability-based security
  approach to manage access control in the {Internet} of things,''
  \emph{Mathematical and Computer Modelling}, vol.~58, pp. 1189--1205, 2013.

\bibitem{bhatt2017access}
S.~Bhatt, F.~Patwa, and R.~Sandhu, ``Access control model for {AWS} {Internet}
  of things,'' in \emph{Proc. International Conference on Network and System
  Security}, 2017, pp. 721--736.

\bibitem{gusmeroli2012iot}
S.~Gusmeroli, S.~Piccione, and D.~Rotondi, ``{IoT} access control issues: A
  capability based approach,'' in \emph{Proc. 2012 Sixth International
  Conference on Innovative Mobile and Internet Services in Ubiquitous
  Computing}, 2012, pp. 787--792.

\bibitem{liu2012authentication}
J.~Liu, Y.~Xiao, and C.~P. Chen, ``Authentication and access control in the
  {Internet} of things,'' in \emph{Proc. 2012 32nd International Conference on
  Distributed Computing Systems Workshops}, 2012, pp. 588--592.

\bibitem{ouaddah2017access}
A.~Ouaddah, H.~Mousannif, A.~A. Elkalam, and A.~A. Ouahman, ``Access control in
  the {Internet} of things: Big challenges and new opportunities,''
  \emph{Computer Networks}, vol. 112, pp. 237--262, 2017.

\bibitem{weber2010internet}
R.~H. Weber, ``Internet of things--{New} security and privacy challenges,''
  \emph{Computer Law \& Security Review}, vol.~26, pp. 23--30, 2010.

\bibitem{pilkington2016blockchain}
M.~Pilkington, ``Blockchain technology: Principles and applications,'' in
  \emph{Research handbook on digital transformations}.\hskip 1em plus 0.5em
  minus 0.4em\relax UK: Edward Elgar Publishing, 2016.

\bibitem{mam}
\BIBentryALTinterwordspacing
Introducing masked authenticated messaging--{IOTA}. [Online]. Available:
  \url{https://blog.iota.org/introducing-masked-authenticated-messaging-e55c1822d50e/.}
\BIBentrySTDinterwordspacing

\bibitem{mainnet}
\BIBentryALTinterwordspacing
{IOTA} networks-- {IOTA} documentation. [Online]. Available:
  \url{https://docs.iota.org/docs/getting-started/1.1/networks/overview.}
\BIBentrySTDinterwordspacing

\bibitem{mamapi}
\BIBentryALTinterwordspacing
Masked authentication messaging wrapper for {Javascript} (browser and node).
  [Online]. Available: \url{https://github.com/iotaledger/mam.client.js/.}
\BIBentrySTDinterwordspacing

\bibitem{zlwen}
\BIBentryALTinterwordspacing
zlwen/cpabe-java: The implementation of ciphertext policy attribute based
  encryption in {Java}. [Online]. Available:
  \url{https://github.com/zlwen/cpabe-java/.}
\BIBentrySTDinterwordspacing

\bibitem{iotatx}
\BIBentryALTinterwordspacing
Transaction fields-- {IOTA} documentation. [Online]. Available:
  \url{https://docs.iota.org/docs/getting-started/1.1/references/transaction-fields.}
\BIBentrySTDinterwordspacing

\bibitem{iotapow}
\BIBentryALTinterwordspacing
Sending transactions-- {IOTA} documentation. [Online]. Available:
  \url{https://docs.iota.org/docs/getting-started/1.1/first-steps/sending-transactions.}
\BIBentrySTDinterwordspacing

\end{thebibliography}
\vspace{12pt}

\end{document}